\documentclass{article}

\usepackage{PRIMEarxiv}
\usepackage{authblk}
\usepackage{float}
\usepackage{tabularx}
\usepackage[table]{xcolor}
\usepackage{listings}

\usepackage{amsmath}
\usepackage[utf8]{inputenc} 
\usepackage[T1]{fontenc}    
\usepackage{hyperref}       
\usepackage{url}            
\usepackage{acronym}
\usepackage{booktabs}       
\usepackage{amsfonts}       
\usepackage{nicefrac}       
\usepackage{microtype}      
\usepackage{lipsum}
\usepackage{fancyhdr}       
\usepackage{graphicx}       
\graphicspath{{media/}}     

\title{Automation-Exploit: A Multi-Agent LLM Framework for Adaptive Offensive Security with Digital Twin-Based Risk-Mitigated Exploitation
\thanks{\textit{\underline{Preprint Notice}}: \textbf{This manuscript is under review at Computers \& Security (Elsevier).}} 
}

\author{Biagio Andreucci}
\author{Arcangelo Castiglione}
\affil{Dipartimento di Informatica, Università degli Studi di Salerno, Fisciano (SA), 84084, Italy}
\affil[ ]{\texttt{biagioo2001.73@gmail.com, acastiglione@unisa.it}}

\begin{document}
\maketitle

\begin{abstract}
\noindent\rule{\linewidth}{0.4pt} 
\vspace{0.2cm}

\begin{quote} 
    The offensive security landscape is highly fragmented: enterprise platforms avoid memory-corruption vulnerabilities due to \ac{dos} risks, \ac{aeg} systems suffer from semantic blindness, and \ac{llm} agents face safety alignment filters and ``Live Fire'' execution hazards. We introduce \textit{Automation-Exploit}, a fully autonomous \ac{mas} framework designed for adaptive offensive security in complex Black-Box scenarios. It bridges the abstraction gap between reconnaissance and exploitation, by autonomously exfiltrating executables and contextual intelligence across multiple protocols, utilizing this data to fuel both logical and binary attack chains. Crucially, the framework introduces an adaptive safety architecture to mitigate \ac{dos} risks. While it natively resolves logical and web-based vulnerabilities, it employs a conditional isomorphic validation for high-risk memory-corruption flaws: provided the target binary is successfully exfiltrated, it dynamically instantiates a cross-platform Digital Twin. By enforcing strict state synchronization, including \textit{libc} alignment and runtime File Descriptor Hooking, potentially destructive payloads are iteratively debugged in an isolated replica. This enables a highly risk-mitigated, ``One-Shot'' execution on the physical target. Empirical evaluations across eight scenarios, including undocumented Zero-Day environments to rule out \ac{llm} Data Contamination, validate the framework's architectural resilience, demonstrating its ability to prevent ``Live Fire'' crashes and execute risk-mitigated compromises on actual targets.

    \vspace{0.4cm}
    \textbf{Keywords:} Offensive Security $\cdot$ Large Language Models $\cdot$ Multi-Agent Systems $\cdot$ Digital Twin $\cdot$ Automatic Exploit Generation $\cdot$ Autonomous Penetration Testing
\end{quote}

\vspace{0.1cm}
\noindent\rule{\linewidth}{0.4pt} 
\end{abstract}

\clearpage 

\section*{Nomenclature}
\renewcommand{\baselinestretch}{0.75}\normalsize
\renewcommand{\aclabelfont}[1]{\textsc{\acsfont{#1}}}
\begin{acronym}[CAPEC]
\acro{llm}[LLM]{Large Language Model}
\acro{mas}[MAS]{Multi-Agent System}
\acro{dos}[DoS]{Denial of Service}
\acro{aeg}[AEG]{Automatic Exploit Generation}
\acro{roi}[ROI]{Return on Investment}
\acro{ctem}[CTEM]{Continuous Threat Exposure Management}
\acro{va}[VA]{Vulnerability Assessment}
\acro{pt}[PT]{Penetration Testing}
\acro{cve}[CVE]{Common Vulnerabilities and Exposures}
\acro{cpe}[CPE]{Common Platform Enumeration}
\acro{nvd}[NVD]{National Vulnerability Database}
\acro{cvss}[CVSS]{Common Vulnerability Scoring System}
\acro{epss}[EPSS]{Exploit Prediction Scoring System}
\acro{cot}[CoT]{Chain-of-Thought}
\acro{icl}[ICL]{In-Context Learning}
\acro{rlhf}[RLHF]{Reinforcement Learning from Human Feedback}
\acro{tag}[TAG]{Tool-Augmented Generation}
\acro{rag}[RAG]{Retrieval-Augmented Generation}
\acro{api}[API]{Application Programming Interface}
\acro{capec}[CAPEC]{Common Attack Pattern Enumeration and Classification}
\acro{osint}[OSINT]{Open-Source Intelligence}
\acro{waf}[WAF]{Web Application Firewall}
\acro{ips}[IPS]{Intrusion Prevention System}
\acro{apt}[APT]{Advanced Persistent Threat}
\acro{rce}[RCE]{Remote Code Execution}
\acro{aslr}[ASLR]{Address Space Layout Randomization}
\acro{pie}[PIE]{Position Independent Executable}
\acro{rop}[ROP]{Return-Oriented Programming}
\acro{uaf}[UAF]{Use-After-Free}
\acro{fnse}[FNSE]{False Negative Syntax Errors}
\acro{ger}[GER]{Global Efficiency Ratio}
\acro{ttc}[TTC]{Time-to-Compromise}
\acro{fpr}[FPR]{False Positive Rate}
\acro{aer}[AER]{Action Executability Rate}
\acro{hitl}[HITL]{Human-in-the-Loop}
\acro{fd}[FD]{File Descriptor}
\acro{jit}[JIT]{Just-In-Time}
\acro{edr}[EDR]{Endpoint Detection and Response}
\acro{gpu}[GPU]{Graphics Processing Unit}
\end{acronym}
\renewcommand{\baselinestretch}{1}\normalsize

\clearpage 

\raggedbottom 
\section{Introduction}
\label{sec:intro}

The offensive security landscape is undergoing a fundamental transformation as organizations increasingly recognize that proactive vulnerability assessment is essential to defending against sophisticated cyber threats. Traditional security testing methodologies, while valuable, have historically been constrained by a fundamental trade-off between operational safety and exploitation depth. On one hand, deterministic vulnerability scanners can rapidly enumerate potential weaknesses across large networks but suffer from semantic blindness, generating numerous false positives while lacking the capability to validate actual exploitability [15, 16]. On the other hand, manual penetration testing conducted by human experts achieves remarkable depth and contextual understanding but remains prohibitively slow, expensive, and difficult to scale across modern, dynamic infrastructures.

In response to these limitations, the research community has pursued three parallel trajectories toward automation. First, enterprise Continuous Threat Exposure Management (CTEM) platforms such as Pentera [17], NodeZero [18], and XBOW [19] have excelled at semantic scavenging and lateral movement, autonomously navigating network environments to validate exposure paths. However, these platforms operate under a strict ``Safe Exploitation'' paradigm, utilizing benign payloads, such as simple DNS queries, to theoretically prove access without altering target state [25]. To avert Denial of Service (DoS) risks on critical infrastructure, they deliberately halt at memory-corruption vulnerabilities (e.g., buffer overflows), lacking the isolated infrastructure required to iteratively test and guarantee exploit stability.

Second, Automatic Exploit Generation (AEG) systems, exemplified by symbolic execution engines like Mayhem [26], S2E [27], and angr [28], have demonstrated mathematical precision in exploring memory paths to autonomously forge shellcode for complex vulnerabilities. Yet these technologies suffer from profound semantic blindness: they cannot navigate web applications, interpret contextual hints, or autonomously apply extraction techniques to exfiltrate target executables in black-box scenarios. Consequently, they invariably require human operators to manually supply isolated binaries, precluding fully autonomous, end-to-end offensive operations.

Third, the emergence of Large Language Models has introduced the possibility of combining investigative intuition with dynamic code generation. Early frameworks like PentestGPT [15] demonstrated significant reasoning capabilities, while recent autonomous architectures including CHECKMATE [29], PwnGPT [16], and PentestAgent [31] have achieved fully autonomous operation. Despite their cognitive potential, state-of-the-art LLM agents face two fundamental barriers. First, reliance on commercial cloud APIs subjects them to strict safety alignment (RLHF), causing models to frequently refuse exploit generation due to embedded ethical policies [9]. Second, lacking isolated simulation infrastructures, these agents routinely execute generated code directly on physical targets. While this unstructured ``Live Fire'' approach may tolerate basic web injections, attempting complex memory-corruption exploits exposes infrastructure to unacceptable risks of unintended crashes and DoS.

The offensive security ecosystem thus lacks a critical ``logical bridge'' connecting the semantic richness of web reconnaissance with the mathematical precision required for binary exploitation. Enterprise platforms avoid memory corruption for safety, AEG systems suffer from semantic blindness, and current LLM agents falter against both ethical constraints and live-fire execution hazards. This research gap is further compounded by the absence of any framework capable of autonomously extracting target binaries from black-box environments, instantiating isomorphic replicas for safe testing, and delivering validated exploits in a single, risk-mitigated execution.

To address these deficiencies, this paper introduces Automation-Exploit, a fully autonomous Multi-Agent System (MAS) framework designed for complex black-box scenarios. The framework bridges the ``abstraction leap'' between reconnaissance and exploitation by autonomously exfiltrating executables and contextual intelligence across multiple protocols, to autonomously fuel complex logical and binary attack chains. Crucially, to mitigate DoS risks, the framework introduces conditional isomorphic validation: upon successful exfiltration of a target binary, it dynamically instantiates a cross-platform Digital Twin. By enforcing extreme synchronization, including libc alignment and runtime File Descriptor Hooking, destructive payloads are iteratively debugged in an isolated replica, guaranteeing risk-mitigated, ``One-Shot'' execution on the actual target. The framework further overcomes intrinsic LLM limitations through an Adversarial Hand-off mechanism that bypasses cloud API safety alignment, an Adaptive Pruning mechanism via a Navigator agent that dynamically truncates fruitless attack vectors, and a Two-Stage Adversarial Auditing protocol that significantly reduces false positives from generative hallucinations. Evaluations across eight scenarios, including undocumented zero-day environments to rule out LLM data contamination, demonstrate the framework's robust autonomous reasoning, polymorphic flexibility in constructing Digital Twins (Linux, Windows Native, WINE), and ability to prevent ``Live Fire'' crashes and execute risk-mitigated compromises on actual targets.

\subsection{Contributions of the Paper}
\label{subsec:contributions}
This research advances autonomous offensive security by overcoming the deterministic limits of traditional scanners and the operational hazards of generative agents. The architecture introduces three core pillars that ensure autonomy, efficiency, and execution safety:

\begin{itemize}
    \item \textbf{Bridging the Semantic-Mathematical Gap:} \textit{Automation-Exploit} resolves the historical divide between contextual web reconnaissance and the mathematical precision required for binary exploitation. By autonomously exfiltrating target assets and contextual intelligence (e.g., configuration files, source code) across multiple protocols, it extracts critical data to fuel complex vulnerability chaining across distinct services.
    \item \textbf{``Risk-Mitigated'' Execution via Isomorphic Digital Twins:} While the use of Digital Twins and cyber ranges for security testing is an established practice \cite{cyberrange2020}, our primary novelty lies in integrating this paradigm directly into an autonomous Multi-Agent LLM architecture specifically tailored for binary exploitation. Unlike systems that risk \ac{dos} with unstructured ``Live Fire'' executions, the framework solves the efficacy-safety dilemma through a predictive Safety Layer \cite{digitaltwin2023}. By dynamically instantiating a cross-platform replica and enforcing extreme synchronization, including \texttt{libc} alignment and runtime File Descriptor Hooking, LLM agents can iteratively debug destructive payloads in an isolated sandbox. This enables a highly risk-mitigated, ``One-Shot'' execution on actual targets.
    \item \textbf{Structural Mitigation of \ac{llm} Limitations:} The system actively overcomes intrinsic generative model defects. It integrates a novel ``Adversarial Hand-off'' to bypass Cloud \ac{api} Safety Alignment, an ``Adaptive Pruning'' mechanism via a Navigator agent to dynamically truncate fruitless attack vectors, and a ``Two-Stage Adversarial Auditing'' protocol that cross-references forensic logs to significantly mitigate generative hallucinations discussed in \cite{llmhallucination2025}.
\end{itemize}

\subsection{Paper Organization}
\label{subsec:organization}
The remainder of this paper is organized as follows. Section \ref{sec:background} provides the foundational background on offensive security, Large Language Models, and Digital Twin technologies. Section \ref{sec:sota} critically reviews the state of the art, defining the specific research gap addressed by our work. Section \ref{sec:architecture} presents the logical architecture and the proposed Multi-Agent System design, while Section \ref{sec:implementation} details its technical implementation. Section \ref{sec:evaluation} discusses the experimental evaluation, highlighting the framework's efficiency, cognitive adaptability, and safety guarantees. Section \ref{sec:ethics} addresses the ethical considerations inherent to offensive security automation, including responsible disclosure and dual-use implications. Finally, Section \ref{sec:conclusions} outlines the conclusions and directions for future work.

\section{Background}
\label{sec:background}
This section reviews four foundations of Automation-Exploit: offensive security fundamentals, Large Language Models for security automation, autonomous and multi-agent systems, and Digital Twin and virtualization technologies. Together, they define the technological context and motivate our architectural choices.

\subsection{Offensive Security Fundamentals}
\label{subsec:offensive_sec}
Following NIST SP 800-115 \cite{nist800115}, offensive security distinguishes between breadth-first \ac{va} and depth-first \ac{pt}. While \ac{va} produces unverified vulnerability lists, \ac{pt} actively simulates adversaries to exploit flaws and validate concrete risks. \textit{Automation-Exploit} autonomously executes \ac{pt}, elevating operations beyond passive scanning.

To formalize these adversarial behaviors, the framework maps its operations to the MITRE ATT\&CK matrix \cite{mitreattack}, independently navigating critical phases: Reconnaissance (TA0043), Initial Access (TA0001), and Execution (TA0002) \cite{mitreattack}. Effective automation requires standardized taxonomies, utilizing \ac{cve} identifiers and the \ac{cpe} schema to map network assets to public flaws \cite{nistcpe}. However, since static \ac{cvss} scores often fail to reflect real-world exploitability \cite{epss2021}, the framework integrates dynamic risk metrics like the \ac{epss} to prioritize attacks based on actual likelihood rather than theoretical severity \cite{epss2021}.

\subsection{Large Language Models in Security Automation}
\label{subsec:llm_security}
While \ac{llm}s demonstrate advanced logical reasoning, their probabilistic nature inherently introduces syntactic and semantic hallucinations, necessitating robust validation architectures \cite{llmhallucination2025}. To optimize inferential accuracy for multi-step vulnerability analysis, techniques such as \ac{cot} and \ac{icl} are routinely employed \cite{cotreasoning2022}. Furthermore, deploying commercial \ac{llm}s in offensive contexts encounters strict safety alignments (\ac{rlhf}). \textit{Automation-Exploit} addresses these operational constraints through an Adversarial Hand-off mechanism, decoupling semantic planning from code generation to maintain operational continuity \cite{jailbreak2023}.

\subsection{Autonomous Agents and Multi-Agent Systems}
\label{subsec:mas}
Shifting from monolithic models to Cognitive Agents \cite{llmagentsurvey2024} enables dynamic environment interaction. In offensive security, this is realized through \ac{tag} to execute deterministic external tools \cite{toolformer2023} and \ac{rag} to ground prompts in authoritative vulnerability data \cite{selfrag2023}. Our framework adopts a \ac{mas} architecture, decomposing the exploitation lifecycle into specialized ``Personas'' (e.g., \textit{Drafter}, \textit{Reviewer}). This segregation strictly bounds contextual visibility, a design choice we implemented to prevent the cognitive overload \cite{contextbleeding2024} and the representation drift documented in \cite{wei2025shadows}.

\subsection{Digital Twin and Virtualization Technologies}
\label{subsec:digital_twin}
To safely evaluate inherently destructive exploits (e.g., memory corruption) on critical infrastructures, execution must be decoupled from live production environments \cite{digitaltwin2023, cyberrange2020}. Building upon these foundational virtualization concepts, our framework introduces a predictive Safety Layer that actively interfaces with generative agents. By enforcing extreme isomorphic alignment — dynamically synchronizing 
shared libraries and hooking runtime file descriptors — the isolated 
replica perfectly mirrors the target's networking and execution 
behavior \cite{cyberrange2020}.

\section{State of the Art and Research Gap}
\label{sec:sota}
This section reviews existing approaches to vulnerability detection, exploit generation, and LLM-driven penetration testing. It examines vulnerability scanners, CTEM platforms, academic AEG systems, LLM-based offensive agents, and tool-augmented open-source frameworks. Our analysis identifies a key gap: no current approach combines web reconnaissance, binary exploitation, and risk-mitigated live-fire execution autonomously. We then show how Automation-Exploit addresses this gap.

\subsection{Deterministic Vulnerability Scanners and Enterprise Platforms}
\label{subsec:deterministic_scanners}
Historically, deterministic scanners have dominated vulnerability detection using rigid pattern-matching against static databases. Despite rapid scalability, they suffer from intrinsic semantic blindness \cite{pentestgpt2024, pwngpt2025}, generating numerous false positives and lacking active exploitation capabilities, shifting the verification burden entirely onto human operators. 

To overcome these limits, \ac{ctem} platforms like Pentera \cite{pentera} and NodeZero \cite{nodezero} excel in semantic scavenging and lateral movement. Recently, the commercial landscape has expanded to include AI-driven and exposure management platforms such as XBOW \cite{xbow2026}, PlexTrac \cite{plextrac2026}, XM Cyber \cite{xmcyber2026}, as well as established ecosystem modules from Tenable \cite{tenablevm2023}, Qualys \cite{moncy2023qualys}, and Wiz \cite{wizcnapp2026}. 

However, they all operate under a strict ``Safe Exploitation'' paradigm, utilizing benign payloads (e.g., simple DNS queries) to theoretically prove access without altering the target's state \cite{pentera_safe}. Consequently, to avert Denial of Service (\ac{dos}) risks on critical infrastructure, these enterprise platforms deliberately halt at memory-corruption vulnerabilities (e.g., Buffer Overflows) on custom binaries. Lacking an isolated infrastructure to iteratively test and guarantee exploit stability, they strictly avoid binary exploration in production environments.

\subsection{Automatic Exploit Generation Systems}
\label{subsec:aeg}
Parallel to network scanners, academic research advanced binary exploitation via \ac{aeg} systems. Symbolic execution engines like Mayhem \cite{mayhem}, S2E \cite{s2e}, and angr \cite{angr} excel at mathematically exploring memory paths to autonomously forge shellcode for complex vulnerabilities. However, these technologies suffer from profound ``Semantic Blindness''. Despite their mathematical precision, \ac{aeg} engines cannot comprehend the broader infrastructural context. They are inherently incapable of navigating web applications, interpreting contextual hints, or autonomously applying exfiltration techniques to extract target executables in Black-Box scenarios. Consequently, they invariably require human operators to manually supply the isolated binary, precluding fully autonomous, end-to-end offensive operations.

\subsection{LLM-Based Offensive Agents}
\label{subsec:llm_agents}
\ac{llm}s introduce the possibility of combining investigative intuition with dynamic code generation. Early frameworks like PentestGPT \cite{pentestgpt2024} demonstrated significant reasoning but operated in a \ac{hitl} modality, restricting end-to-end autonomy. To overcome this, recent frameworks, including CHECKMATE \cite{wang2025}, PwnGPT \cite{pwngpt2025}, AutoPentest \cite{henke2025}, and PentestAgent \cite{pentestagent2025}, have implemented fully autonomous architectures. 

Despite their cognitive potential, state-of-the-art \ac{llm} agents face two fundamental barriers. First, relying on Cloud \ac{api}s subjects them to strict Safety Alignment (\ac{rlhf}), causing models to frequently refuse exploit generation due to embedded ethical policies \cite{jailbreak2023}. Second, lacking isolated simulation infrastructures, these agents routinely execute generated code directly on physical targets. While this unstructured ``Live Fire'' approach may tolerate basic web injections, attempting complex memory-corruption exploits exposes the infrastructure to unacceptable risks of unintended crashes and \ac{dos}. This instability renders current autonomous agents unsuitable for evaluating mission-critical production environments.

\subsection{Open-Source Frameworks and Tool-Augmented Agents}
\label{subsec:opensource_agents}
Recently, the open-source community has introduced several \ac{llm}-driven orchestrators and projects, such as PentAGI \cite{pentagi2026}, Strix \cite{strix2026}, Deadend CLI \cite{deadendcli2026}, CAI (Cyber AI) \cite{caicyberai2026}, and MCP-Security modules \cite{mcpsecurity2026}, designed to integrate autonomous agents with traditional security tools. While these tool-augmented systems effectively automate repetitive tasks and generate proof-of-concept exploits within isolated sandboxes, they predominantly focus on web and logical vulnerabilities. They lack the architectural depth required for complex multi-agent meta-reasoning and the extreme isomorphic synchronization (e.g., Digital Twin) needed to safely evaluate and transition memory-corruption exploits to physical targets without causing system instability.

\subsection{Critical Comparative Analysis}
To better delineate the research gap addressed by \textit{Automation-Exploit}, it is essential to contrast its architecture with recent autonomous agents such as \textit{CHECKMATE} \cite{wang2025}, \textit{PwnGPT} \cite{pwngpt2025}, and \textit{PentestAgent} \cite{pentestagent2025}, as well as open-source tool-augmented projects \cite{pentagi2026, strix2026, deadendcli2026}.

\paragraph{Evasion of Safety Filters}
While \textit{PentestAgent} and \textit{CHECKMATE} primarily rely on prompt engineering or semantic reframing to navigate Cloud API safety alignments (RLHF) \cite{wang2025, pentestagent2025}, \textit{Automation-Exploit} introduces a structural decoupling via the \textbf{Adversarial Hand-off}. Unlike existing pipelines that may face intermittent refusals when generating complex exploitation logic, our framework utilizes a locally-hosted model to bootstrap a raw technical skeleton. This local stage effectively shifts the cloud model's task to ``Code Repair'', leveraging semantic masking techniques \cite{semanticmask2025} to address deterministic guardrails without compromising reasoning depth.

\paragraph{Strategic Planning vs. Adaptive Pruning}
The decision-making core of \textit{CHECKMATE} is grounded in \textbf{classical planning} to ensure logical consistency \cite{wang2025}. In contrast, the \textbf{Navigator} agent in \textit{Automation-Exploit} adopts an \textbf{Adaptive Pruning} mechanism. While classical planning excels in deterministic environments, it often struggles with the ``environmental friction'' found in Black-Box scenarios. The Navigator performs retrospective analysis on forensic logs to issue \texttt{ABORT} or \texttt{JUMP} verdicts, allowing the framework to truncate fruitless attack vectors more dynamically than rigid pre-calculated planners.

\paragraph{Open-Source Wrappers vs. MAS Segregation}
Recently, the open-source community has introduced several LLM-driven orchestrators, such as \textit{PentAGI} \cite{pentagi2026}, \textit{Strix} \cite{strix2026}, and \textit{Deadend CLI} \cite{deadendcli2026}, designed to integrate autonomous agents with traditional security tools. However, these tools predominantly function as single-agent wrappers. They pipe monolithic LLM outputs directly into standard offensive tools in a linear sequence, making them highly susceptible to context window saturation and generative hallucinations \cite{contextbleeding2024}. \textit{Automation-Exploit} fundamentally diverges by enforcing strict Multi-Agent System (MAS) segregation. By dividing the workload among specialized Personas, the framework heavily bounds the context window, preventing representation drift.

\paragraph{Binary Exploitation and Digital Twins}
Finally, whereas \textit{PwnGPT} focuses extensively on LLM-based Automatic Exploit Generation (AEG) for binaries, it lacks the end-to-end web reconnaissance capabilities to exfiltrate those binaries autonomously \cite{pwngpt2025}. \textit{Automation-Exploit} bridges this ``abstraction leap'' by autonomously extracting hidden assets. Furthermore, unlike open-source alternatives that are structurally restricted to non-state-altering web exploits, and other agents that execute code in a ``Live Fire'' mode directly on the target, our framework integrates a \textbf{Digital Twin} for isomorphic validation \cite{digitaltwin2023}. This ensures that high-risk memory-corruption vulnerabilities, which are typically avoided by platforms like \textit{Pentera} or \textit{NodeZero} due to DoS risks \cite{pentera_safe}, can be safely debugged before a single ``One-Shot'' delivery on the physical target.

\subsection{Research Gap and Architectural Comparison}
\label{subsec:research_gap}
The offensive security ecosystem lacks a ``logical bridge'' connecting semantic web reconnaissance with the mathematical precision of binary exploitation. Enterprise platforms avoid the latter for safety, \ac{aeg} systems suffer from semantic blindness, and current \ac{llm} agents falter against ethical constraints and ``Live Fire'' execution hazards.

\textit{Automation-Exploit} resolves these deficiencies. First, it bridges the abstraction gap by autonomously exfiltrating executables and contextual intelligence across multiple protocols. This data is injected into the cognitive loop, enabling autonomous vulnerability chaining where artifacts exfiltrated from one service are correlated to compromise adjacent services operating on different ports. Second, it resolves the ``Safe Exploitation'' versus ``Live Fire'' dichotomy via a conditional Safety Layer. Upon exfiltrating a target's executable, it generates an isomorphic Digital Twin. Through \textit{libc} synchronization and File Descriptor hooking, the agent tests destructive payloads in isolated replicas, enabling a highly risk-mitigated, single-attempt ``One-Shot'' execution on actual targets.

Furthermore, an ``Adversarial Hand-off'' decouples semantic planning from code generation to address Cloud \ac{api} Safety Alignment constraints \cite{jailbreak2023}, while a Two-Stage Adversarial Auditing protocol \cite{llmhallucination2025} employs an independent reviewer to scrutinize forensic logs and reject generative hallucinations. 

\begin{table}[H]
\rowcolors{2}{gray!10}{white} 
\centering
\caption{Architectural Comparison of Automation-Exploit against State-of-the-Art Paradigms.}
\label{tab:sota_comparison}
\footnotesize 
\renewcommand{\arraystretch}{1.8} %
\begin{tabularx}{\textwidth}{@{} 
    >{\raggedright\arraybackslash\hsize=1.1\hsize}X 
    >{\raggedright\arraybackslash\hsize=1.0\hsize}X 
    >{\raggedright\arraybackslash\hsize=1.0\hsize}X 
    >{\raggedright\arraybackslash\hsize=1.1\hsize}X 
    >{\raggedright\arraybackslash\hsize=0.8\hsize}X 
    @{}}
\toprule
\rowcolor{gray!30}
\textbf{Feature} & \textbf{Automation-Exploit} & \textbf{CTEM Platforms} & \textbf{AEG Systems} & \textbf{Autonomous LLMs} \\
\rowcolor{gray!30}
& \scriptsize\textit{(Proposed)} & \scriptsize\textit{(e.g., Pentera, NodeZero)} & \scriptsize\textit{(e.g., Mayhem, angr)} & \scriptsize\textit{(e.g., PentestGPT)} \\ 
\midrule

\textbf{Binary Exploit Handling} & \textbf{Yes.} Digital Twin ensures reliability before ``One-Shot'' execution. & \textbf{No.} Avoids DoS-inducing memory exploits. & \textbf{Partial.} Requires user-provided binaries (Semantic Blindness). & \textbf{Dangerous.} ``Live Fire'' mode risks severe DoS. \\ 

\textbf{Execution Safety (DoS Mitigation)} & \textbf{``Risk-Mitigated''.} Iterative debugging via synchronized Digital Twin (libc, \acs{fd}). & \textbf{``Safe Exploitation''.} Uses benign, non-state-altering payloads. & \textbf{N/A.} Executes in emulated analysis environments. & \textbf{High Risk.} No simulation layer; payloads hit production directly. \\ 

\textbf{End-to-End Automation} & \textbf{Yes.} Fully autonomous (Autonomous exfiltration to forensic validation). & \textbf{Partial.} Discovery and non-destructive lateral movement only. & \textbf{No.} Requires manual target supply and interpretation. & \textbf{Partial.} Often stops at web/application layers. \\ 

\textbf{Alignment Constraint Handling} & \textbf{``Adversarial Hand-off''.} Local models draft logic; cloud models refactor. & \textbf{N/A.} Deterministic; no public LLMs used. & \textbf{N/A.} Math-based deterministic tools. & \textbf{Problematic.} Cloud APIs cause frequent refusals. \\ 

\textbf{Result Validation (Anti-Hallucination)} & \textbf{Two-Stage Auditing.} Independent agent cross-references forensic logs. & \textbf{Deterministic.} Checks benign payload execution. & \textbf{Mathematical.} Exploit generation is a demonstrable path. & \textbf{Weak.} Output parsing leads to high false positives. \\ 

\textbf{Strategic Adaptability} & \textbf{Yes.} Navigator agent dynamically prunes fruitless attack branches. & \textbf{Limited.} Static prioritization (e.g., EPSS); mostly linear. & \textbf{N/A.} Exhaustive, non-adaptive path exploration. & \textbf{Linear.} Static plans with poor runtime meta-reasoning. \\ 
\bottomrule
\end{tabularx}
\end{table}

Finally, it overcomes static execution by decoupling static prioritization (Decision Engine) from runtime meta-reasoning. A Navigator agent applies Adaptive Pruning, issuing explicit \texttt{ABORT} or \texttt{JUMP} verdicts to truncate fruitless attack vectors and dynamically reallocate resources.

To systematically position the proposed framework against the aforementioned paradigms, Table \ref{tab:sota_comparison} summarizes the architectural and operational differences across key offensive security capabilities.

\section{Proposed Framework: Architecture and Design}
\label{sec:architecture}
This section presents the architecture of Automation-Exploit, addressing the gap between semantic web reconnaissance and precise binary exploitation under safe operating conditions. It introduces the framework’s key concepts, outlines its multi-agent cognitive architecture and knowledge management, and describes the decision, safety, and validation mechanisms that enable fully autonomous, risk-mitigated offensive security.

By structurally segregating these roles, the framework ensures that the Fixer maintains surgical precision without the entropic overload of obsolete global strategies, while the Reviewer significantly mitigates the generative \ac{llm}'s self-delusion tendencies \cite{multiagentdebate2023}. We hypothesize that this strict role segregation between semantic reasoning and code generation systematically mitigates context window saturation \cite{contextbleeding2024} and prevents representation drift \cite{wei2025shadows}. This architectural claim is empirically validated in Section \ref{sec:evaluation}, where the framework maintains a highly stable \ac{aer} between 85\% and 100\% across complex scenarios.

\subsection{Multi-Agent Cognitive Architecture}
\label{subsec:mas_architecture}
\textit{Automation-Exploit} employs a modular \ac{mas} architecture to overcome the severe cognitive bottlenecks of monolithic \ac{llm}s. As context windows expand with dense technical data, monolithic models suffer from ``Lost in the Middle'' phenomena \cite{contextbleeding2024,li2024long} and Representation Drift \cite{wei2025shadows}, which degrade reasoning accuracy and drastically increase hallucinations. 

To systematically mitigate this, the framework decouples the offensive workload into specialized sub-agents (Personas). Each agent is confined by a strictly bounded System Prompt, ensuring it processes only the specific inputs required for its task. Table \ref{tab:ai_personas} summarizes the cognitive segregation, defining the roles, I/O flows, and operational constraints of the core agents.

\begin{table*}[htbp]
\centering
\caption{Taxonomy of AI Personas in the \textit{Automation-Exploit} Framework.}
\label{tab:ai_personas}
\resizebox{\textwidth}{!}{%
\renewcommand{\arraystretch}{1.4}
\rowcolors{2}{gray!10}{white}
\begin{tabular}{@{}p{3cm}p{4.5cm}p{3.5cm}p{3.5cm}p{4.5cm}@{}}
\toprule
\rowcolor{gray!30}
\textbf{Agent Persona} & \textbf{System Prompt Role} & \textbf{Primary Inputs} & \textbf{Outputs / Verdict} & \textbf{Operational Constraints} \\ \midrule

\textbf{The Recon Hunter} & Autonomous Asset Hunter \& Data Miner & Target IP/Port, \texttt{HTTP} headers, previously exfiltrated assets history. & Python crawling and exfiltration scripts. & Enforces heuristic boundary constraints to prevent infinite recursion and manages stream-based artifact downloading. \\ 

\textbf{Objective Setter} & Security Strategist & \ac{cve} metadata, vulnerability summaries, attack patterns. & Deterministic Success Marker (e.g., cryptographic execution flags). & Must output exactly one word from the framework's strict deterministic taxonomy. \\ 

\textbf{The Drafter} & Cybersecurity Analyst \& Exploit Developer & Accumulator's strategy, raw \ac{cve} data, target architecture. & Raw exploit code skeleton. & Executed via a locally-hosted \ac{llm} to ensure absolute bypass of initial Cloud-based safety alignments. \\ 

\textbf{The Accumulator} & Lead Security Researcher & Forensic blueprints, past execution logs, target raw data. & Technical specifications / Attack strategy. & Cannot write raw executable code. Invoked exclusively at the start (Iteration 1) of each mini-stage to establish strategic directives. \\ 

\textbf{The Fixer} & Debugging \& Code Repair Specialist & Last executed code, \texttt{stderr}/\texttt{stdout} from sandbox, crash autopsies. & Patched executable payload. & Context strictly truncated to the current mini-stage (Amnesia logic). Focuses purely on code generation following the Accumulator's strategy. \\ 

\textbf{The Navigator} & Senior Red Team Lead \& Efficiency Auditor & Cross-\ac{cve} execution history, routing errors, future task queue. & \texttt{CONTINUE}, \texttt{JUMP}, \texttt{ABORT} & Invoked only at strategic checkpoints to govern resource allocation via Adaptive Pruning. \\ 

\textbf{The Sim Engineers} & Senior Forensic Analyst / Sysadmin & Exfiltrated target binaries, raw intelligence data, magic bytes. & OS signatures, Docker configs, Registry scripts. & Operates strictly on the Digital Twin logic. Focuses purely on building isolated, isomorphic replicas. \\ 

\textbf{The Sanitizer} & Code Sanitizer \& Linter & Raw \ac{llm} generation containing markdown fences or conversational artifacts. & Clean, strictly executable code. & Powered by lightweight models (e.g., Gemini 2.5 Flash) for high-speed, low-cost text stripping prior to execution. \\ 

\textbf{The Reviewer} & Merciless Security Auditor & Executed code, forensic logs, Digital Twin autopsy data. & Structured Verdict (e.g., Verified / False Positive). & Read-only; cannot modify code. Enforces strict forensic causality (e.g., Differential Access checks). \\ \bottomrule
\end{tabular}%
}
\end{table*}

\textbf{Adaptive Model Allocation and Cognitive Scaling.} 
To optimize the computational budget and mitigate API constraints, Automation-Exploit eschews static model assignments in favor of a dynamic, risk-aware allocation strategy. The framework selectively provisions Large Language Models (LLMs) based on the cognitive complexity of the task at hand. Initial payload bootstrapping is strictly delegated to the \textit{Drafter} agent, powered by a locally hosted, uncensored model (e.g., Mistral 7B). This localization is structurally imperative to execute the ``Adversarial Hand-off'' mechanism, inherently bypassing the stringent safety alignments \cite{jailbreak2023} enforced by commercial Cloud APIs. Once the initial refusal barrier is circumvented, routine and high-frequency cognitive workloads, such as semantic refinement, syntax linting by the \textit{Fixer}, simulation engineering, and forensic auditing by the \textit{Reviewer}, are allocated to a standard frontier model (e.g., Gemini 2.5 Pro). 

Conversely, tasks demanding advanced meta-reasoning, such as the Adaptive Pruning executed by the \textit{Navigator}, are elevated to a high-cognition tier (e.g., Gemini 3 Pro). Crucially, the framework enforces a systemic ``Cognitive Upgrade'' specifically during \textit{Clone Mode}. When operating within the fully isolated confines of the Digital Twin, all generative tasks are promoted to the highest available tier. In this highly controlled state, exhausting the inferential budget to iteratively solve complex memory-corruption vulnerabilities is strategically justified, as the absolute isolation neutralizes \ac{dos} risks to the actual target. 

To further optimize operational efficiency, the framework isolates purely mechanical, low-cognition tasks from the primary inferential engines. For instance, the \textit{Sanitizer} agent leverages high-speed, cost-effective models (e.g., Gemini 2.5 Flash) to rapidly strip conversational artifacts and markdown fences prior to execution. Finally, to guarantee continuity against network latency, API rate-limiting, or server timeouts, the Orchestrator integrates a \textit{Smart Downgrade} fallback mechanism. This cognitive circuit breaker dynamically degrades the inferential tier (e.g., cascading from Gemini 3 Pro back to Gemini 2.5 Pro) following consecutive failures, thereby preserving the exploitation pipeline's integrity without halting the execution.

\subsection{Unified Knowledge Base for Context Persistence}
\label{subsec:knowledge_base}
To bridge the semantic gap that traditionally creates vertical technological silos, the \textit{Automation-Exploit} framework centralizes all gathered intelligence into a persistent state repository, which serves as a dynamic Unified Knowledge Base. To prevent Large Language Model context window saturation and the consequent degradation in reasoning accuracy during long-horizon operations \cite{contextbleeding2024, li2024long}, the system enforces strict context persistence by structuring its operational memory within this directory using lightweight, machine-readable structured data formats. 

This logical architecture systematically aggregates a vast array of target-related data, including, but not limited to, detailed port enumerations, vulnerability metadata, and the physical artifacts (e.g., configuration files and source code) exfiltrated during the reconnaissance phase. Crucially, during the exploitation phase, this centralized repository allows the sequential cognitive loop to transcend individual port silos. Agents can draw upon critical intelligence or binaries extracted from one port to fuel complex attack chains or instantiate Digital Twins for vulnerabilities discovered on entirely different ports. By maintaining this persistent and universally accessible state, the framework autonomously correlates cross-port data to compromise adjacent services without suffering from representation drift \cite{contextbleeding2024, wei2025shadows}.

\textbf{Sequential Access and Hash-Based Data Hygiene.} In the current implementation, the multi-agent cognitive loop operates in a strictly sequential pipeline orchestrated by the lifecycle manager. This architectural design inherently prevents race conditions, ensuring that sub-agents do not concurrently overwrite the same contextual state. To maintain structural hygiene and prevent the proliferation of redundant artifacts during the exfiltration phase, the framework incorporates an active deduplication mechanism. Before any extracted asset or telemetry data is recorded into the structured operational memory, its cryptographic checksum is computed and cross-referenced against the existing database. Duplicate files are autonomously purged, ensuring the knowledge base remains lean and preventing the LLM from processing redundant tokens. 

Furthermore, to guarantee resilience against state serialization corruption (e.g., resulting from abrupt container terminations or runtime I/O interruptions), the orchestrator strictly enforces schema validation before any state update. By relying on atomic sequential writes and rigorous hash-based deduplication, the framework ensures that the cognitive loop is never poisoned by malformed or duplicate context data.

\subsection{Decision Engine and Adaptive Pruning}
The Decision Engine orchestrates the Automation-Exploit framework to avoid static, brute-force execution by formalizing vulnerability data into a structured queue of actionable tasks. Formally, a computational \textit{Task} ($\mathcal{T}$) is defined as a tuple:

\begin{equation}
    \mathcal{T} = \langle ID, S, V, M_{env}, P \rangle
\end{equation}

where $ID$ is a unique identifier, $S$ denotes the target service configuration (IP, Port, Protocol), $V$ encapsulates vulnerability metadata (e.g., CVE, \acs{capec} ID, or AI-synthesized logic), $M_{env}$ represents the set of exfiltrated environmental artifacts (e.g., binaries, configuration files from the exfiltration phase), and $P$ is the priority score.

To prioritize execution based on actual operational \ac{roi}, the framework replaces static metrics with a composite $priority\_score \in [0, 100]$. This is formally calculated as:

\begin{equation}
    P = \alpha \cdot EPSS + \beta \cdot CVSS_{base} + \gamma \cdot f_{context}(M_{env})
\end{equation}

where $EPSS$ represents the dynamic probability of real-world exploitation \cite{epss2021}, $CVSS_{base}$ provides the foundational severity, and $f_{context}$ is a heuristic function that rewards the presence of physical artifacts (e.g., successful exfiltration of the target executable), drastically elevating the task's ROI. All three components are normalized to $[0,1]$ prior to scoring: EPSS is
natively bounded in $[0,1]$~\cite{epss2021}; \textit{CVSS}$_{\text{base}}$
is divided by~10; and $f_{\text{context}}(M_{\text{env}})$ is computed as
$\min\!\left(1,\, S_{\text{bonus}} / S_{\text{max}}\right)$, where
$S_{\text{bonus}}$ is the cumulative heuristic score from the context
weighting table (cf.\ Table~2) and $S_{\text{max}} = 950$ is the maximum
attainable bonus across all tactical conditions.
The coefficients are heuristically assigned as $\alpha = 0.6$,
$\beta = 0.2$, $\gamma = 0.2$ (with $\alpha + \beta + \gamma = 1$),
reflecting the EPSS-first design rationale: dynamic exploitability
is weighted threefold relative to either static severity or
environmental context.

To manage this queue during runtime, the framework employs the Navigator agent's Adaptive Pruning mechanism. Invoked after exhausting the maximum iteration threshold ($\tau_{max}$) for a specific \ac{cve} or Task, the Navigator performs a retrospective analysis of execution logs and runtime errors \cite{reflexion2023, dataforensics2006}. Based on this review, it issues one of three binding routing verdicts: \texttt{CONTINUE} (proceed nominally), \texttt{JUMP} (skip unviable vectors due to technological mismatch), or \texttt{ABORT} (truncate the entire attack branch and pivot to the next enumerated port).

\subsection{Safety Layer: Digital Twin-Based Protection}
\label{subsec:safety_layer}
To mitigate \ac{dos} risks on physical infrastructures, a conditional Safety Layer employs the Digital Twin paradigm \cite{digitaltwin2023}. If a target executable is successfully exfiltrated during the reconnaissance phase, the system dynamically instantiates a cross-platform, containerized replica. The core hypothesis driving this design is that ``Live Fire'' execution of memory-corruption exploits by autonomous agents is structurally unsafe~\cite{pentestgpt2024,cyberrange2020}. By enforcing an isomorphic replica, we postulate that critical \ac{dos} conditions can be safely absorbed---a claim directly corroborated by our empirical results in Section \ref{sec:evaluation}, where the Digital Twin successfully intercepted 14 critical \ac{dos} conditions across Scenarios G and H.

Before authorizing cognitive debugging, the framework validates the replica's fidelity via a formal \textbf{Behavioral Mirror Test}. Let $\mathcal{P}$ be a predefined set of active network probes (e.g., null bytes, malformed HTTP headers) and $R(p)$ denote the deterministic response vector (e.g., banner strings, socket persistence state, error codes) for a given probe $p \in \mathcal{P}$. The Digital Twin is certified as operationally isomorphic if and only if it satisfies strict behavioral symmetry with the physical target:

\begin{equation}
    \forall p \in \mathcal{P}, \quad R_{target}(p) \equiv R_{twin}(p) \pm \Delta t
\end{equation}

where $\Delta t$ accounts for acceptable environmental network latency. 

Once isomorphism is formally certified, to guarantee flawless exploit translation to the actual target, the framework enforces extreme synchronization. This includes exact \texttt{libc} alignment and runtime File Descriptor Hooking (via \texttt{LD\_PRELOAD} and \texttt{dup2}) \cite{kerrisk2010} to perfectly replicate the host's socket I/O architecture \cite{cyberrange2020}. 

Cognitive agents iteratively debug destructive payloads against this isolated clone, which features a Self-Healing mechanism for instantaneous crash restoration \cite{cyberrange2020}. Upon crashes, an Autopsy mechanism extracts critical execution feedback, enabling the cognitive loop to reason about potential causes and refine the payload in subsequent iterations.

\subsection{Two-Stage Validation Protocol}
\label{subsec:validation_protocol}
A major challenge with autonomous language agents is their susceptibility to generative hallucinations and self-delusion, often leading to premature declarations of successful compromise \cite{llmhallucination2025}. To combat this verification unreliability, \textit{Automation-Exploit} concludes its execution cycle with a Two-Stage Validation Protocol:

\begin{itemize}
    \item \textbf{Stage 1: Execution and Flagging.} The operative sub-agents (Drafter and Fixer) execute the payload. Perceived successes are temporarily flagged for review rather than immediately accepted.
    \item \textbf{Stage 2: Adversarial Auditing.} An independent Reviewer agent intervenes as an objective auditor. Without generating code, it exclusively cross-references the executed payload against correlated forensic logs, standard outputs, and system errors.
\end{itemize}

By objectively confirming compromises through this multi-agent adversarial evaluation, the Reviewer significantly mitigates the self-delusion tendencies of generative models \cite{multiagentdebate2023}, ensuring a highly improved reliability of reported vulnerabilities.

\section{Implementation}
\label{sec:implementation}
Translating the architectural principles of Section \ref{sec:architecture} into a deployable system requires resolving non-trivial engineering challenges: deterministic protocol identification \cite{nmapbook}, context saturation under iterative LLM inference \cite{contextbleeding2024}, and isomorphic fidelity in cross-platform emulation \cite{cyberrange2020}. The following subsections detail how each structural challenge is resolved against contemporary alternatives.

\subsection{Reconnaissance and Asset Scavenging}
\label{subsec:reconnaissance}
To address the semantic misclassification problem identified in §\ref{sec:sota} --- where passive enumeration fails to resolve services on non-standard ports \cite{nmapbook} --- this framework adopts the principle of active Protocol Bifurcation, which we hypothesize reduces reconnaissance noise sufficiently to anchor downstream vulnerability mapping; this is empirically supported by the 77.2\% operational time distribution reported in Section~\ref{sec:evaluation}. Concretely, a Discovery Engine performs global port discovery and resolves protocol ambiguity via active HTTP \texttt{HEAD} interrogation, replacing passive banner grabbing with deterministic service fingerprinting.

Post-enumeration, normalized data is prioritized by the Decision Engine, triggering the exfiltration phase for sequential asset extraction. Employing techniques like Heuristic Bounded Extraction, agents autonomously extract sensitive artifacts (e.g., binaries, configurations) into an isolated digital drop-zone. Prior to cognitive loop injection, a data hygiene procedure applies cryptographic hashing to purge duplicates, centralizing only unique intelligence within the persistent state repository.

\subsection{Vulnerability Intelligence and Task Generation}
\label{subsec:vuln_intel}
To address the gap between raw enumeration output and actionable exploitation targets --- a limitation shared by both deterministic scanners and heuristic agents reviewed in Section~\ref{sec:sota} --- this framework adopts the principle of heterogeneous task unification, hypothesizing that normalizing CVE-based and AI-inferred vectors into a single prioritized schema maximizes exploitable ROI \cite{epss2021}; this is empirically validated by the GER results across all eight scenarios in Section~\ref{subsec:quantitative}. The Vulnerability Intelligence module implements this principle via a dual-engine process: deterministic mapping converts enumeration data into \ac{cpe} v2.3 format \cite{nistcpe} to query the \acs{nvd}, enriched with \acs{capec} \cite{mitrecapec} and MITRE ATT\&CK \cite{mitreattack} context.

To address undocumented logical flaws, a Heuristic Tasking Engine introduces inferential AI-Tasking. This heuristic engine analyzes topology, assigning high interest scores to sensitive endpoints (e.g., \texttt{/admin}, \texttt{/login}) or exposed banners to synthesize custom attack vectors. Finally, both deterministic \ac{cve}s and AI-Tasks are normalized into a unified structured schema with a normalized priority metric, enabling the Decision Engine to evaluate heterogeneous vectors based on actual \ac{roi} \cite{epss2021}.

\subsection{Hybrid LLM Orchestration}
\label{subsec:llm_orchestration}
\begin{figure}[H]
    \centering
    \includegraphics[width=0.75\textwidth]{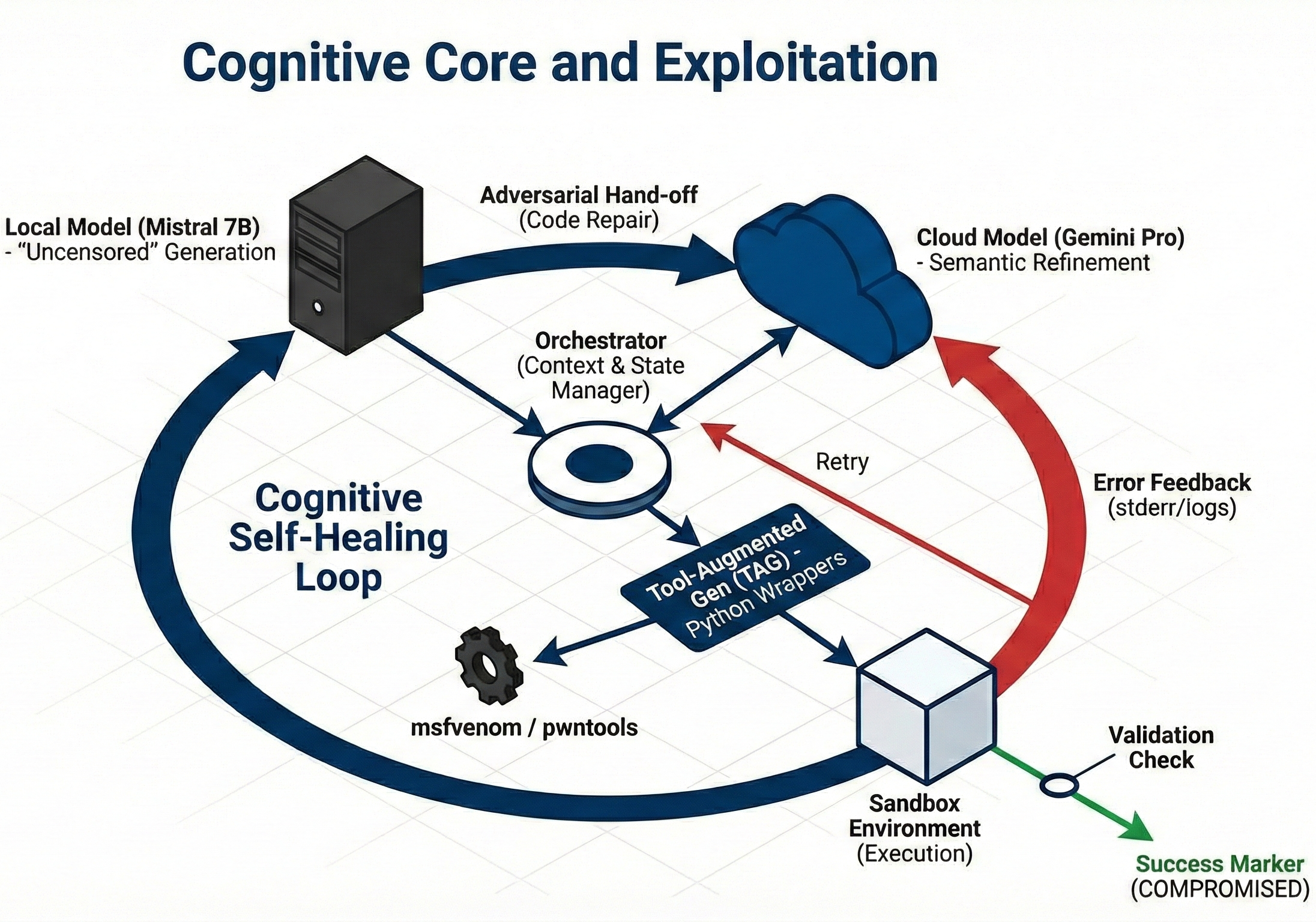} 
    \caption{\textbf{Cognitive Self-Healing Loop.} Flow of the ``Adversarial Hand-off'' between local and cloud models, mediated by the Orchestrator and \ac{tag}.}
    \label{fig:cognitive_loop}
\end{figure}

To address the operational constraint imposed by Safety Alignment in commercial Cloud models \cite{jailbreak2023} --- identified in Section~\ref{sec:sota} as a primary barrier to autonomous exploit generation --- this framework adopts the principle of generative decoupling, hypothesizing that separating semantic reasoning from code synthesis neutralizes alignment-induced refusals without degrading output quality; this is empirically validated by the longitudinal resilience results in Section~\ref{subsec:quantitative}. The Hybrid Orchestration module instantiates this principle via an ``Adversarial Hand-off'': a local Mistral 7B model bootstraps raw exploit skeletons, which Gemini Pro subsequently refines via Semantic Masking reframed as ``Code Repair'' \cite{semanticmask2025}. This orchestration and its feedback loop are illustrated in Fig.~\ref{fig:cognitive_loop}.

To maintain resilience, an external payload repository allows the dynamic injection of \textit{System Instruction Overrides} (e.g., adversarial poetry, Base64, or low-resource languages) directly into the model's system prompt without altering core logic \cite{advpoetry2025, yuan2023cipherchat, yong2023lowresource}.

\textbf{Cognitive Checkpoints and Progressive Context Injection.} 
To strictly operationalize the ``abstraction leap'', the \textit{Orchestrator} implements progressive Cognitive Checkpoints within the self-healing loop to systematically prevent context window saturation and representation drift \cite{contextbleeding2024, wei2025shadows}. Rather than blindly iterating payload syntax, the framework enforces a ``Tactical Pause'' exactly at Iteration 7-B. During this Double-Shot Forensics phase, the agent temporarily halts blind exploit generation to execute an analytical probe against the previously exfiltrated assets. By extracting contextual intelligence, such as hardcoded credentials or hidden directory structures, and re-injecting it into the context window, the framework performs a semantic pivot, autonomously bridging the gap between passive reconnaissance and active exploitation. 

Furthermore, if an attack path stagnates at Iteration 10, the system activates a Confidence-Aware Retrieval mechanism. It selectively injects expert external logic (specifically, Metasploit snippets) into the prompt, ranked by semantic relevance to the current attack context. This targeted injection prevents context poisoning \cite{iclpoison2024} while seamlessly guiding the generative model toward proven exploitation primitives via retrieval-augmented generation \cite{selfrag2023}.
\ac{api} stability is governed by the Lifecycle Manager via a Circuit Breaker pattern and Adaptive Model Cascading, downgrading inference under rate-limiting to preserve continuity and \ac{roi} \cite{llmcascading2026}.
Since the local model is utilized exclusively as an ``uncensored semantic Trojan horse'' to bypass initial refusal policies, its isolated reasoning performance is intentionally not evaluated. Its sole architectural purpose is to generate a rough, even if logically flawed, tactical skeleton, deliberately delegating all semantic refinement and logical debugging to the frontier Cloud model.

\subsection{Digital Twin and Forensic Pipeline}
\label{subsec:forensic_pipeline}
To address the DoS hazard inherent in iterative exploit refinement on physical targets --- the core limitation of ``Live Fire'' paradigms identified in Section~\ref{sec:sota} \cite{digitaltwin2023} --- this framework adopts the principle of isomorphic pre-validation, hypothesizing that destructive payloads can be safely stabilized on a structurally equivalent replica before One-Shot execution on the actual target; this is empirically validated by the crash absorption data in Section~\ref{subsec:quantitative} (Table~\ref{tab:crashes_prevented}). The Isomorphic Instantiator implements this principle by dynamically provisioning multi-platform containers (Linux, Windows Native, WINE) directly from exfiltrated artifacts, autonomously inferring the target OS via binary signature analysis and switching the host hypervisor accordingly.\footnote{Platform detection is performed via Magic Byte inspection (e.g., \texttt{MZ} for PE binaries, \texttt{0x7FELF} for ELF), enabling zero-configuration cross-platform provisioning.} Before exploitation, an internal behavioral mirror test strictly validates the replica's isomorphism.

\begin{figure}[H]
    \centering
    \includegraphics[width=\textwidth]{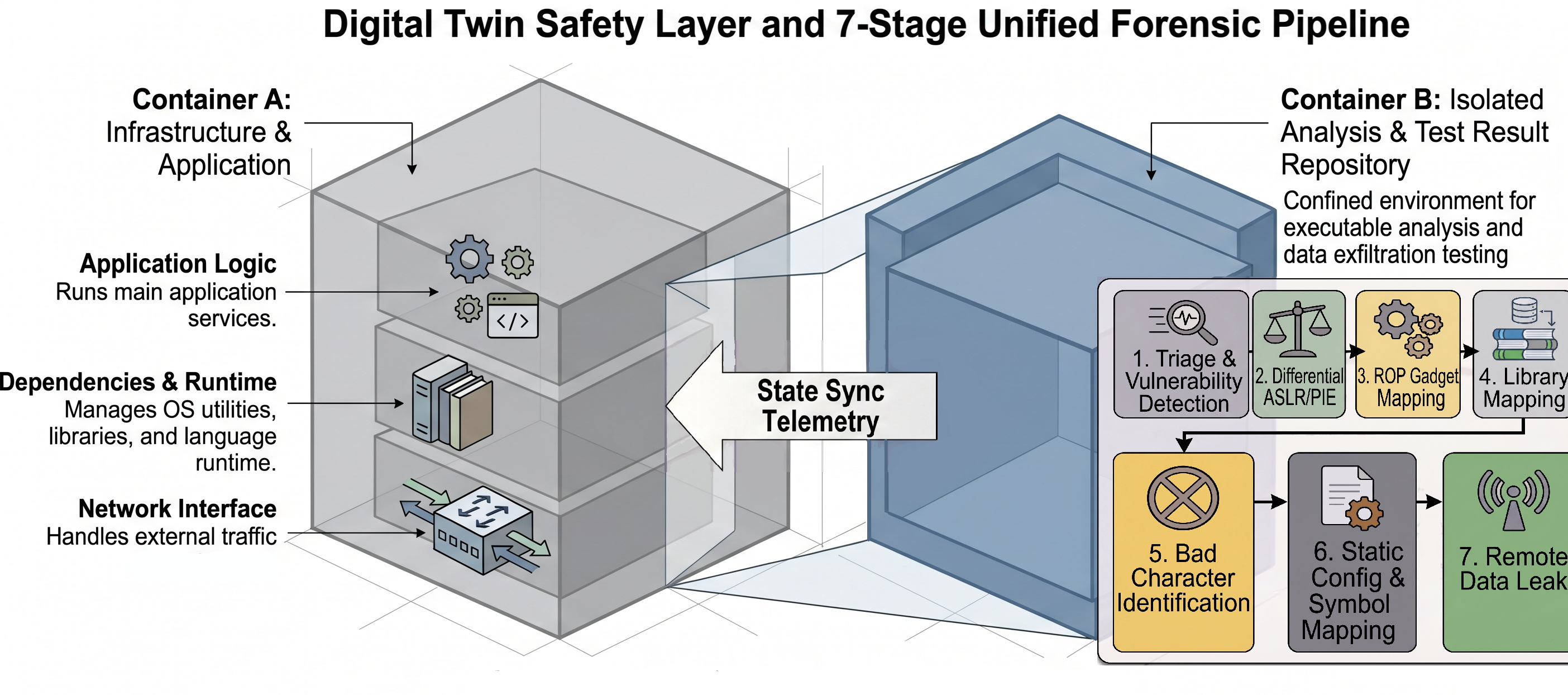} 
    \caption{\textbf{Digital Twin Safety Layer and Unified Forensic Engine.} Container A maintains isomorphic fidelity with the actual target via state synchronization, while Container B safely executes a 7-stage forensic pipeline to extract vulnerability blueprints without risking \ac{dos} on the replica.}
    \label{fig:digital_twin_architecture}
\end{figure}

To extract a deterministic \textit{Ground Truth} without polluting the replica, the framework deploys a decoupled \textit{Unified Forensic Engine}. As illustrated in Figure \ref{fig:digital_twin_architecture}, it executes a 7-stage analytical pipeline inspired by \ac{aeg} methodologies \cite{pwngpt2025}, incorporating automated static reverse engineering of the exfiltrated executable alongside dynamic fuzzing for EIP/RIP offsets, \acs{aslr}/\acs{pie} detection, and ROP gadget extraction. 

An \acs{aslr}-Aware sanitizer formats this telemetry into a structured vulnerability blueprint to anchor the \ac{llm}'s context.

To ensure a flawless transition to the actual target, the framework enforces network topology mirroring via runtime File Descriptor synchronization, guaranteeing that exploits validated on the Twin are structurally portable to the physical target without modification.\footnote{Concretely, an ephemeral C shared object injected via \texttt{LD\_PRELOAD} hooks the \texttt{accept()} syscall and invokes \texttt{dup2} to dynamically override local file descriptors, mirroring the target's network topology \cite{kerrisk2010}.}

\begin{figure}[H]
    \centering
    \includegraphics[width=\textwidth]{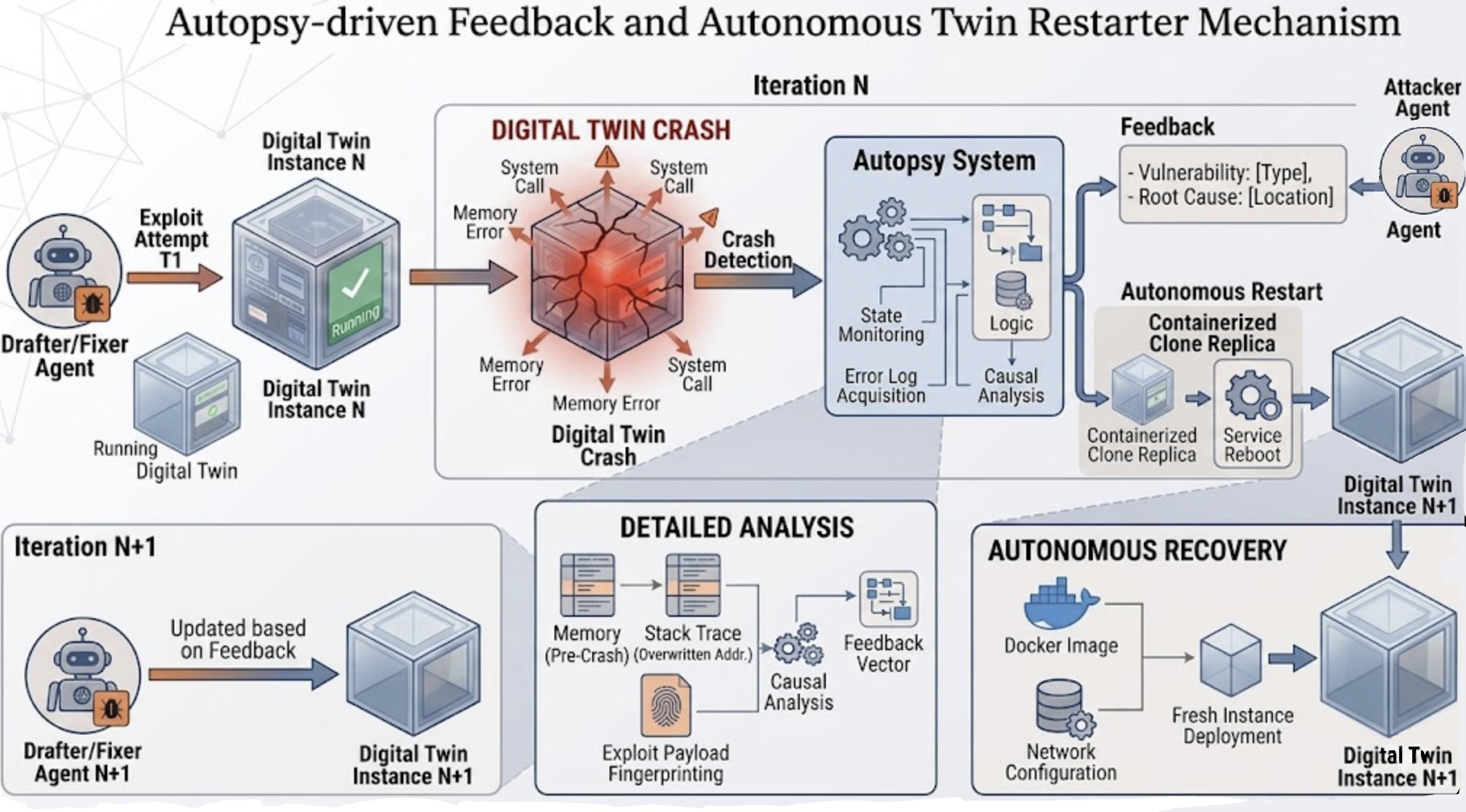}
    \caption{\textbf{Autopsy-driven Feedback and Self-Healing Mechanism.} When an unstable payload triggers a critical failure within the Digital Twin, the Autopsy System intercepts the crash telemetry. This deterministic feedback is processed to autonomously deploy a fresh replica and provide register-level intelligence to the generative agents for the subsequent iteration.}
    \label{fig:autopsy_log}
\end{figure}

Finally, an asynchronous Autopsy Callback guarantees operational resilience. Upon a crash, it extracts Core Dumps or Windows MiniDumps, feeding explicit register states back to the \ac{llm} for deterministic Self-Healing \cite{cyberrange2020}. As detailed in the architectural workflow in Figure \ref{fig:autopsy_log}, when a fatal payload triggers a service crash on the Digital Twin, the framework autonomously extracts the exact register state (eax, ebx, esp, etc.). It then securely re-provisions a fresh, uncontaminated replica (Twin Instance N+1) while simultaneously providing the \ac{llm} with the necessary deterministic telemetry to refine the payload and achieve isomorphic stability in the subsequent iteration.

\subsection{Execution Sandbox and Validation}
\label{subsec:execution_sandbox}
To address the structural fragility of LLM-generated code in heterogeneous runtime environments --- a failure mode documented in \cite{llmhallucination2025, wei2025shadows} and identified in Section~\ref{sec:sota} as a key driver of low executability rates in existing agents --- this framework adopts the principle of adaptive execution isolation, hypothesizing that runtime language detection combined with semantic output sanitization maximizes Action Executability Rate without human intervention; this is empirically validated by the AER results ($\geq$85\% in five of eight scenarios) reported in Section~\ref{subsec:quantitative}. The Execution Engine implements this principle via a \ac{jit} polyglot pipeline within an isolated virtualized sandbox, monitored by a Sentinel Health Check to detect library corruption \cite{cyberrange2020}. This dynamic adaptation is a deliberate design choice required to maintain high Action Executability Rates (\ac{aer}), ensuring that generative outputs are correctly compiled (e.g., C code) or natively interpreted without requiring manual intervention. To minimize \ac{fnse}, a dedicated text sanitizer aggressively strips Markdown fences and metadata from \ac{llm} outputs.

To combat generative hallucinations documented in \cite{llmhallucination2025}, the framework replaces standard exit-code checks with a Two-Stage Validation Protocol. Stage 1 performs deterministic signal processing on \texttt{stdout} and \texttt{stderr}, verifying success tags, enforcing Differential Access baselines (e.g., HTTP 401/403 before bypass), and applying Anti-Echo filtering. If flagged, Stage 2 invokes an independent \ac{llm} acting as a ``Merciless Security Auditor'' \cite{multiagentdebate2023}. Operating on an Actor-Critic debate model, this Reviewer cross-references the code with forensic logs and Autopsy telemetry. By objectively rejecting logical hallucinations, such as misinterpreting 200 OK responses as successful logins, the Reviewer ensures only verified exploits are reported.

\section{Experimental Evaluation}
\label{sec:evaluation}
To validate the hypotheses embedded in the framework's design---specifically, that multi-agent segregation stabilizes reasoning \cite{multiagentdebate2023} and that a Digital Twin prevents operational \ac{dos} \cite{digitaltwin2023}---this section presents an empirical evaluation across eight scenarios of increasing complexity.

\subsection{Test Methodology and Metrics}
\label{subsec:metrics}

The operational scope of the Automation-Exploit framework is strictly bounded to the Initial Access and Execution phases against a single target IP. The architectural goal is to demonstrate ``depth of exploitation'' (e.g., bridging the gap from web reconnaissance to local binary execution via a Digital Twin) rather than ``breadth of discovery'' or network-wide Lateral Movement (e.g., Active Directory pivoting). Consequently, the experimental scenarios focus on individual, highly complex nodes to isolate and measure the agent's deep reasoning capabilities on a 1-to-1 engagement basis.
\vspace{0.3cm}

\subsubsection{Testing Environment} 
Experiments were conducted on a high-performance mobile workstation equipped with a consumer-grade \ac{gpu} to demonstrate the feasibility of local offloading for the uncensored Mistral 7B (4-bit quantized). This hardware configuration was selected to validate that the framework's local inference requirements are accessible via standard professional-grade laptops, rather than requiring specialized server-side infrastructure. Target environments were isolated within an air-gapped VirtualBox Host-Only network. To evaluate performance, we defined four Key Performance Indicators (KPIs) focusing on quantitative efficiency and cognitive reliability:

\begin{itemize}
    \item \textbf{\ac{ger}:} Quantifies resource optimization by comparing consumed iterations against a theoretical brute-force baseline. It measures the effectiveness of the Decision Engine and Adaptive Pruning. Here, $N_{\text{used}}$ denotes the actual iterations consumed, $N_{\text{potential\_tasks}}$ the number of candidate CVE tasks enqueued by the Decision Engine, and $N_{\text{max\_iter}}$ the maximum iterations allocated per task before it is declared non-exploitable. This ceiling is set to $N_{\text{max\_iter}} = 12$ for standard web and logical CVEs, and extended to $N_{\text{max\_iter}} = 24$ when the Digital Twin Safety Layer is activated for memory-corruption scenarios, to accommodate the iterative Autopsy-driven Self-Healing loop.
    \begin{equation}
        \text{GER} = 1 - \frac{N_{\text{used}}}{N_{\text{potential\_tasks}} \times N_{\text{max\_iter}}}
    \end{equation}
    
    \item \textbf{\ac{ttc}:} Measures absolute operational time from initial reconnaissance to verified payload validation.
    \begin{equation}
        \text{TTC} = t_{\text{end}} - t_{\text{start}}
    \end{equation}

    \item \textbf{False Positive Rate (FPR):} Evaluates the generative model's propensity for logical hallucinations during Stage 1. We deliberately measure the FPR before the Stage 2 Reviewer intervenes. This isolates the intrinsic hallucination rate of the generative LLM and quantifies the ``hidden cost'' of self-delusion: even when a false positive is subsequently neutralized by the Auditor, it represents wasted computational budget and API calls.
    
\begin{equation}
FPR = \frac{N_{FalsePositives}}{N_{TotalIterations}}
\end{equation}

    \item \textbf{\ac{aer}:} Assesses the structural stability of generated code (syntax, runtime, and integration). High \ac{aer} indicates that architectural segmentation successfully mitigates Context Overload and Representation Drift previously defined in \cite{contextbleeding2024, wei2025shadows}.
    \begin{equation}
        \text{AER} = \frac{N_{\text{ValidExecutions}}}{N_{\text{TotalIterations}}}
    \end{equation}
\end{itemize}

\subsection{Case Study Overview}
\label{subsec:case_studies}
To validate architectural autonomy, the framework was tested against eight scenarios of increasing complexity. Initial benchmarks (e.g., Metasploitable, Kioptrix) assessed logical pivoting and adversarial auditing \cite{multiagentdebate2023}, while subsequent tests introduced environmental frictions to trigger Self-Healing and meta-reasoning \cite{reflexion2023}. 

To eliminate Data Contamination bias, where \ac{llm}s retrieve memorized solutions \cite{yang2025, iclpoison2024}, Scenarios G and H utilized custom-built Zero-Day environments. These required the agent to rely exclusively on procedural reasoning and Digital Twin telemetry \cite{digitaltwin2023}. Table \ref{tab:case_studies} summarizes the taxonomy of evaluated scenarios.

\begin{table}[H]
\centering
\small
\renewcommand{\arraystretch}{1.3}
\rowcolors{2}{gray!10}{white} 
\begin{tabular}{@{} l p{2.8cm} l p{7.5cm} @{} }
\hline
\rowcolor{gray!30}
\textbf{ID} & \textbf{Target / Service} & \textbf{Level} & \textbf{Key Capabilities Demonstrated} \\
\hline

\textbf{A} & Metasploitable 2 & Baseline & \textbf{Adversarial Auditing:} Reviewer intercepted FPs, forcing differential access checks. \\

\textbf{B} & Kioptrix 1.1 & Interm. & \textbf{Adaptive Pruning:} Navigator issued \texttt{ABORT} on IPP; pivoted to successful SQLi. \\

\textbf{C} & MrRobot & Advanced & \textbf{Contextual Pivoting:} Synthesized "Lore-Based" dictionaries via pre-trained knowledge. \\

\textbf{D} & DC-2 & Advanced & \textbf{Self-Healing:} Recovered from exceptions; bypassed internal DNS via autonomous scraping. \\

\textbf{E} & Metasploitable 3 & Advanced & \textbf{Meta-Reasoning:} Recalibrated validation logic after evaluating unexpected HTTP responses. \\

\textbf{F} & Brainpan 1 & Advanced & \textbf{Digital Twin (WINE):} Semantic debugging on clones before performing One-Shot \acs{rce}. \\

\textbf{G} & Custom Zero-Day & Extreme & \textbf{Zero Contamination:} Detected \acs{aslr}; autonomously synthesized a Ret2Libc chain via Digital Twin telemetry for \acs{rce}. \\

\textbf{H} & Custom Windows & Extreme & \textbf{Windows Native Twin:} Twin absorbed UAF crashes, enabling deterministic Self-Healing. \\

\hline
\end{tabular}
\vspace{0.2cm}
\caption{Taxonomy of experimental scenarios and validated architectural components.}
\label{tab:case_studies}
\end{table}

\subsection{Quantitative Results and Analysis}
\label{subsec:quantitative}
Quantitative analysis confirms the framework's resource optimization and operational stability across diverse environments.

\subsubsection{Operational Efficiency (GER vs. TTC)}
\begin{figure}[H]
    \centering
    \includegraphics[width=0.75\textwidth]{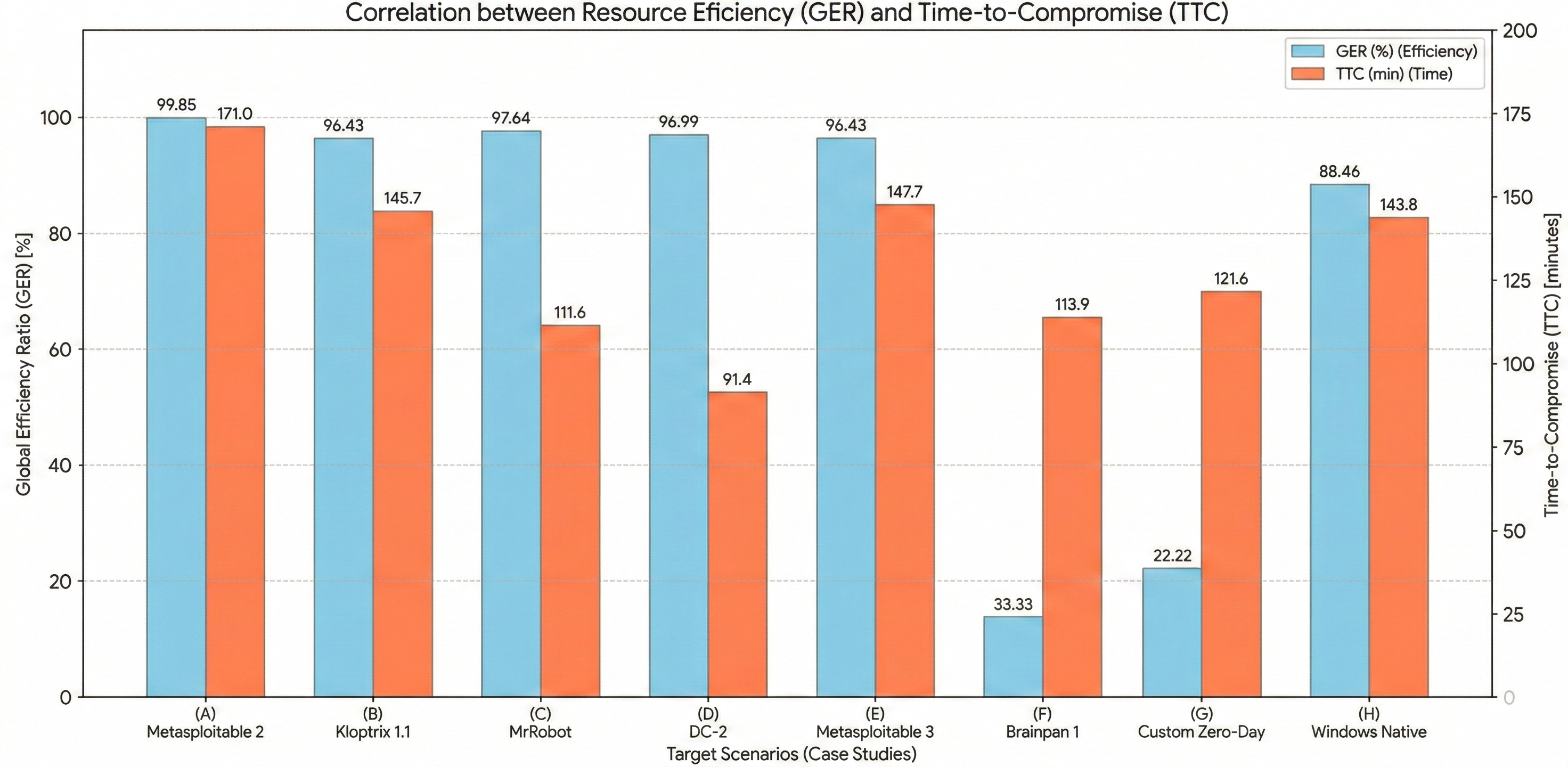}
    \caption{\textbf{Efficiency-Time Correlation.} Global Efficiency Ratio (GER)
    versus Time-to-Compromise (TTC) across all eight scenarios.
    High GER is maintained independently of TTC; drops in Scenarios~F and~G
    reflect constrained binary surfaces requiring exhaustive exploration.}
    \label{fig:ger_ttc_analysis}
\end{figure}

As shown in Fig. \ref{fig:ger_ttc_analysis}, the Global Efficiency Ratio (\ac{ger}) exceeds 96\% in five scenarios (e.g., A: 99.85\%, D: 96.99\%), validating the Navigator's Adaptive Pruning in semantically truncating fruitless ``Rabbit Holes'' \cite{wang2025}. Expected \ac{ger} drops occur in constrained binary targets (F and G) requiring exhaustive exploration of limited surfaces. Conversely, Time-to-Compromise (\ac{ttc}) varies significantly (91.4 mins in D to 171.0 mins in A), lacking linear correlation with efficiency. Extended \ac{ttc}s stem from the cognitive loop's architectural overhead (Digital Twin instantiation, context serialization, \ac{llm} inference) rather than brute-force delays.

\subsubsection{Cognitive Dynamics and Reliability (AER vs. FPR)}
\begin{figure}[H]
    \centering
    \includegraphics[width=0.75\textwidth]{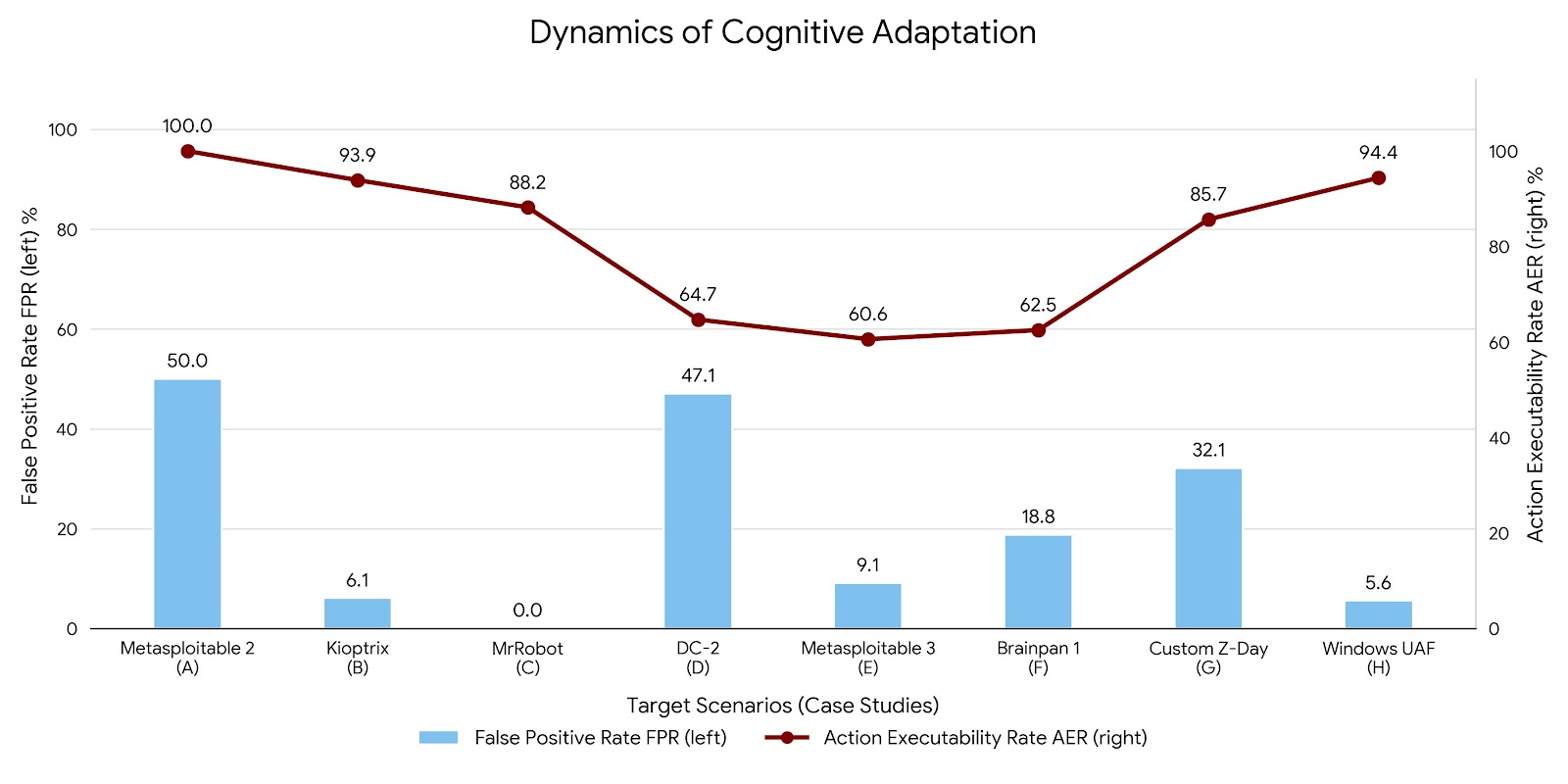}
    \caption{\textbf{Cognitive Adaptation Dynamics.} Action Executability Rate
    (AER) versus False Positive Rate (FPR) during Stage~1 across all eight
    scenarios. AER remains above 85\% in five scenarios; FPR peaks are
    statistical artifacts of low iteration counts and are fully neutralised
    by Stage~2 Adversarial Auditing.}
    \label{fig:adattamento_dinamico}
\end{figure}

Tracking cognitive dynamics (Fig. \ref{fig:adattamento_dinamico}), Action Executability Rate (\ac{aer}) remains highly stable (85\%--100\%) in scenarios A, B, C, G, and H, proving the mini-stage design prevents Information Overload and ensures robust code \cite{contextbleeding2024, wei2025shadows}. \ac{aer} drops (60\%--64\%) in D, E, and F reflect environmental frictions (e.g., host encoding, asynchronous orchestration) rapidly resolved by deterministic Self-Healing \cite{reflexion2023}. \ac{fpr} during Stage 1 remains strictly below 50\%; high peaks (e.g., 50.0\% in Scenario A) are statistical artifacts of low iteration counts. Crucially, the Stage 2 Adversarial Auditing provided a measured, definitive mitigation: across all eight evaluated scenarios, 0 out of the identified Stage 1 false positives were subsequently confirmed as final compromises \cite{multiagentdebate2023}. By rigorously enforcing causal state checks (e.g., Differential Access, Digital Twin logs), the architecture successfully blocked all observed hallucinations from propagating to reported outcomes. Within this experimental sample, no Stage 1 false positive survived the Reviewer's validation, confirming the auditing layer's empirical effectiveness under the tested conditions.

\subsubsection{Temporal Distribution Analysis and Operational Bottlenecks}
To precisely identify the architectural bottlenecks contributing to the extended Time-to-Compromise (TTC) discussed previously, a temporal distribution analysis was conducted across the experimental suite. Interestingly, despite the known architectural complexity of dynamically instantiating an isomorphic replica for security analytics \cite{digitaltwin2023, cyberrange2020}, the operational overhead is demonstrably not dominated by the Safety Layer.

\begin{figure}[H]
    \centering
    \includegraphics[width=0.75\columnwidth]{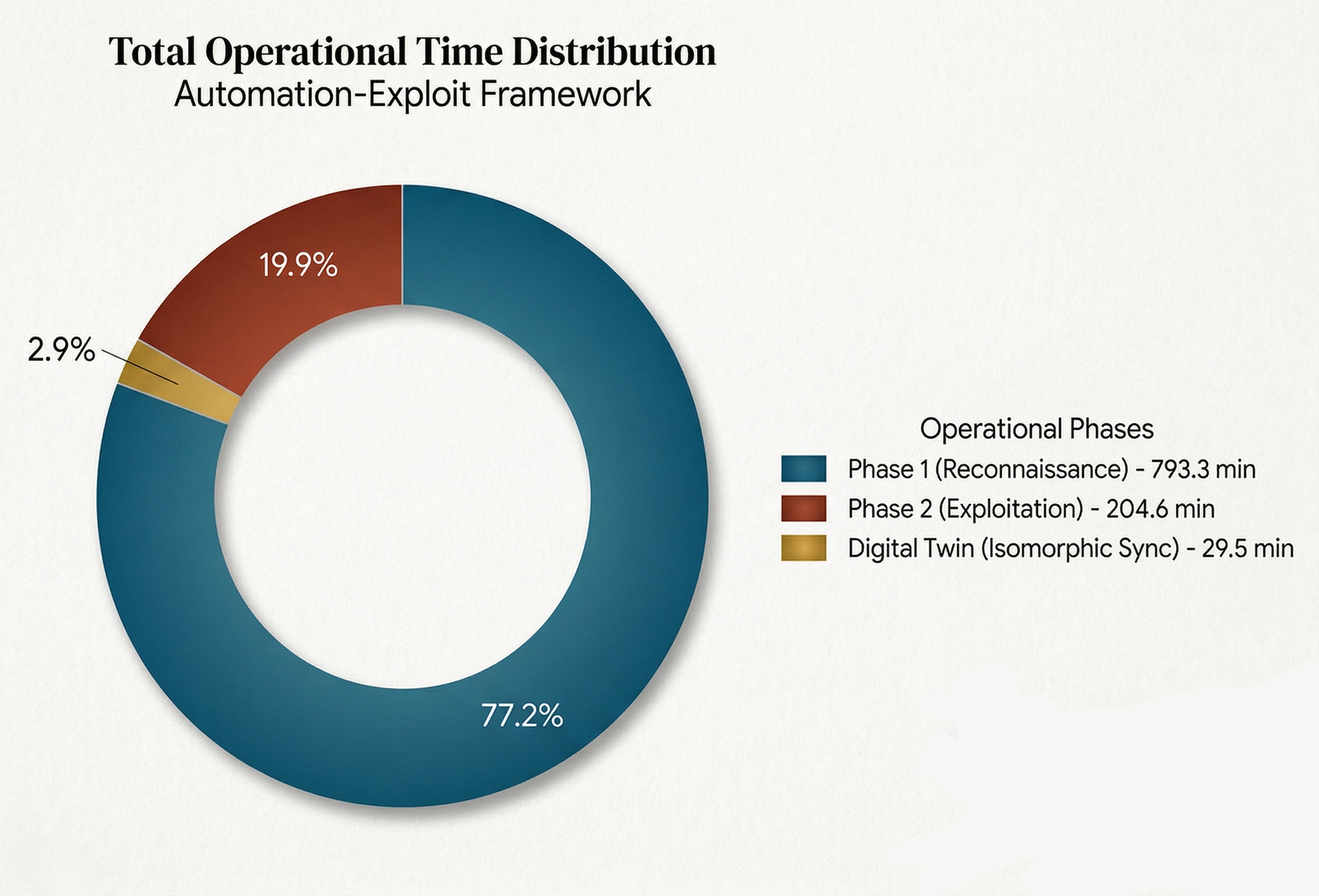} 
    \caption{Total Operational Time Distribution. Phase 1 (Reconnaissance) accounts for 77.2\% of the execution time, while the Digital Twin (Isomorphic Sync) requires only 2.9\%.}
    \label{fig:time_distribution}
\end{figure}

As illustrated in Fig. \ref{fig:time_distribution}, Phase 1 (Reconnaissance and Exfiltration) consumes the vast majority of the operational budget, accounting for 77.2\% (793.3 minutes) of the total time. This bottleneck is directly attributable to the sequential extraction and contextualization of artifacts across multiple identified ports. In contrast, Phase 2 (Exploitation and Cognitive Loop) accounts for 19.9\% (204.6 minutes). 

Crucially, the dynamic instantiation of the Digital Twin (Isomorphic Sync) requires a negligible 2.9\% (29.5 minutes) of the total execution time, primarily because it is a highly optimized process triggered conditionally (e.g., exclusively in memory-corruption scenarios like Scenarios F, G, and H). This empirical data formally validates that ensuring ``Risk-Mitigated'' exploitation via isomorphic replication introduces minimal temporal overhead, isolating the framework's primary bottleneck to sequential I/O reconnaissance, i.e., a limitation explicitly addressed in future asynchronous iterations.

\subsubsection{Empirical Validation of Risk-Mitigated Exploitation}
To quantitatively validate the necessity of the Digital Twin Safety Layer, we tracked the activation frequency of the Autopsy mechanism across the memory-corruption scenarios (F, G, and H). The empirical data unequivocally demonstrates that ``Live Fire'' execution paradigms utilized by state-of-the-art \ac{llm} agents are structurally incompatible with physical infrastructures and actual targets.

As detailed in Table \ref{tab:crashes_prevented}, during the evaluation of the Custom Zero-Day Linux ROP (Scenario G) and the Windows Native \ac{uaf} (Scenario H), the isolated replica safely absorbed a total of 9 and 5 critical \ac{dos} conditions across multiple runs, respectively. Crucially, empirical log analysis reveals that in every successful execution, the deterministic register-level feedback provided by the immediately preceding Autopsy report was the definitive catalyst for achieving a stable payload.

\begin{table}[H]
\centering
\renewcommand{\arraystretch}{1.3}
\rowcolors{2}{gray!10}{white}
\resizebox{\columnwidth}{!}{%
\begin{tabular}{@{} l l c c c p{4.5cm} @{}}
\hline
\rowcolor{gray!30}
\textbf{Scenario} & \textbf{Vulnerability Type} & \textbf{Run 1} & \textbf{Run 2} & \textbf{Run 3} & \textbf{Critical \ac{dos} Conditions Prevented} \\
\hline
\textbf{F (Brainpan)} & Buffer Overflow & 0 & 0 & 0 & \textbf{0} (Offsets extracted natively) \\
\textbf{G (Zero-Day)} & Custom Linux ROP & 1 & 6 & 2 & \textbf{9} \\
\textbf{H (Custom UAF)} & Windows Native \ac{uaf} & 1 & 4 & N/A\textsuperscript{\dag} & \textbf{5} \\
\hline
\end{tabular}%
}
\vspace{0.2cm}
\caption{Critical DoS conditions safely absorbed by the Digital Twin
Safety Layer across memory-corruption scenarios (F, G, H).
$^{\dag}$~Run~3 not conducted: Scenario~H served to validate the
\textit{Adversarial Hand-off} mechanism independently of the
longitudinal test (see Section~5.3); the two completed runs
confirm Digital Twin absorption in every iteration.}
\label{tab:crashes_prevented}
\end{table}

Without the Digital Twin, a single failed iteration would result in an immediate Denial of Service (\ac{dos}) on the physical target. This creates a severe ``Black Box'' dilemma for autonomous agents: if a physical target stops responding, a ``Live Fire'' agent cannot determine whether the connection was dropped by a behavioral-based firewall (NGFW), the payload was neutralized by an Endpoint Detection and Response (EDR) system, or if it caused a catastrophic kernel panic. By strictly confining the iterative failure-and-recovery loop within the isomorphic replica, \textit{Automation-Exploit} resolves this ambiguity. 

Notably, in Scenario F (Brainpan), the Autopsy mechanism recorded zero critical failures. This further validates the framework's forensic efficacy: the Digital Twin allowed the unified forensic engine to safely fuzz and extract the exact Ground Truth offsets prior to execution.

\subsection{Comparative Analysis with State-of-the-Art}
\label{subsec:comparative}
Contrasting \textit{Automation-Exploit} with contemporary paradigms highlights its capacity to bridge existing technological silos:

\textbf{Enterprise \ac{ctem} Platforms (e.g., Pentera, NodeZero, XBOW) \cite{pentera, nodezero, xbow2026}:} To prevent \ac{dos}, commercial platforms avoid memory-corruption vulnerabilities \cite{pentera_safe}. Scenarios F (Brainpan), G (Zero-Day), and H (Windows Native \ac{uaf}) prove our Digital Twin overcomes this by decoupling attacks. It safely absorbs triage crashes, authorizing One-Shot executions only upon validating isomorphic stability.

\textbf{Automatic Exploit Generation (e.g., Mayhem, S2E) \cite{mayhem, s2e}:} While \ac{aeg} engines could theoretically resolve binary Scenarios F, G, and H, their ``Semantic Blindness'' necessitates human-provided executables. Furthermore, being strictly confined to memory exploration, they structurally fail against logical web vulnerabilities (Scenarios A--E), even with human intervention. \textit{Automation-Exploit} transcends these limitations: it autonomously resolves the full web-flaw spectrum (A--E) while matching \ac{aeg} capabilities by autonomously extracting hidden binaries via Heuristic Bounded Extraction for deterministic analysis (F, G, H).

\textbf{Autonomous \ac{llm} and Tool-Augmented Agents (e.g., PentestGPT, CHECKMATE, PentAGI) \cite{pentestgpt2024, wang2025, pentagi2026}:} Cloud-dependent agents face Safety Alignment paralysis \cite{jailbreak2023}, execution loops, and high hallucination rates \cite{llmhallucination2025}. More critically, lacking isolated forensic simulation, their unstructured ``Live Fire'' execution of memory-corruption exploits poses an unacceptable risk of catastrophic \ac{dos} in production environments. 

Rather than relying on traditional success-rate benchmarks that ignore operational safety, we empirically validate this hazard to demonstrate why the ``Live Fire'' paradigm is unsustainable for complex engagements. 

\renewcommand{\lstlistingname}{Log}

\begin{lstlisting}[language=bash,
  float=h,
  basicstyle=\ttfamily\footnotesize, 
  backgroundcolor=\color{gray!10},   
  frame=single,                      
  breaklines=true,
  captionpos=b,                      
  label={fig:crashevidence},
  caption={Empirical Demonstration of Live Fire Hazards. Post-execution \texttt{dmesg} log capturing the critical segmentation fault of the target service (\texttt{auth\_server[1465]}) at \texttt{0xdeadbeef} during Scenario~G.}]
[ 4258.082719] show_signal_msg: 9 callbacks suppressed
[ 4258.082752] auth_server[1465]: segfault at deadbeef ip deadbeef sp bffffd8c 
               error 15
\end{lstlisting}

Log~\ref{fig:crashevidence} documents the kernel-level telemetry (\texttt{dmesg}) captured during Scenario~G. It records a critical segmentation fault (\texttt{segfault at deadbeef ip deadbeef sp bffffd8c error 15}) triggered by an unstable intermediate \ac{rop} payload. This specific experiment empirically demonstrates the hazard of the ``Live Fire'' paradigm: deploying this pre-validation candidate directly on a physical target---as competitor \ac{llm} agents lacking a Digital Twin would do---causes an immediate and irreversible Denial of Service (\ac{dos}).

\textit{Automation-Exploit} systematically prevents such catastrophic failures. In our architecture, the Autopsy module intercepted this flawed payload (containing the invalid return address \texttt{0xdeadbeef}). The resulting crash was safely absorbed by the isolated container, and its forensic telemetry was repurposed to trigger a Self-Healing loop that successfully refined the final exploit chain. Furthermore, the Adversarial Hand-off addresses alignment-induced operational constraints, Adaptive Pruning truncates fruitless ``Rabbit Holes'' (Scenarios C, E), and Two-Stage Adversarial Auditing \cite{multiagentdebate2023} utilizes this exact Autopsy telemetry to definitively reject unstable payloads.

\subsection{Experimental KPI Summary}

Table~\ref{tab:kpi_summary} provides a consolidated overview of the
four key performance indicators measured across all eight experimental
scenarios.

\begin{table}[H]
\centering
\small
\renewcommand{\arraystretch}{1.3}
\rowcolors{2}{gray!10}{white}
\begin{tabular}{@{} l l l r r r r @{}}
\hline
\rowcolor{gray!30}
\textbf{ID} & \textbf{Target} & \textbf{Level} & \textbf{GER (\%)} & \textbf{TTC (min)} & \textbf{FPR (\%)} & \textbf{AER (\%)} \\
\hline

\textbf{A} & Metasploitable 2 & Baseline    & 99.85 & 171.0 & 50.00 & 100.0 \\

\textbf{B} & Kioptrix 1.1     & Interm.     & 96.43 & 145.7 &  6.06 &  93.9 \\

\textbf{C} & MrRobot          & Advanced    & 97.64 & 111.6 &  0.00 &  88.2 \\

\textbf{D} & DC-2             & Advanced    & 96.99 &  91.4 & 47.06 &  64.7 \\

\textbf{E} & Metasploitable 3 & Advanced    & 96.43 & 147.7 &  9.09 &  60.6 \\

\textbf{F} & Brainpan 1       & Advanced    & 33.33 & 113.9 & 18.75 &  62.5 \\

\textbf{G} & Custom Zero-Day  & Extreme     & 22.22 & 121.6 & 32.14 &  85.7 \\

\textbf{H} & Windows Native   & Extreme     & 88.46 & 143.8 &  5.56 &  94.4 \\

\hline
\end{tabular}
\vspace{0.2cm}
\caption{Summary of experimental KPIs across all eight evaluation scenarios. GER: Global Efficiency Ratio; TTC: Time-to-Compromise; FPR: False Positive Rate (generative hallucinations); AER: Action Executability Rate.}
\label{tab:kpi_summary}
\end{table}

\subsection{Discussion of Results and Bias Analysis}
\label{subsec:discussion}
Evaluating \ac{llm}-based agents necessitates addressing Data Contamination bias, whereby models resolve tasks via memorized training data (e.g., public CTF write-ups) rather than authentic deductive reasoning \cite{yang2025, iclpoison2024}. To validate \textit{Automation-Exploit}'s scientific integrity, we critically dissect its performance against this threat:

\begin{itemize}
    \item \textbf{Custom Zero-Days (Definitive Proof):} Scenarios G (Linux Custom Binary) and H (Windows Native \ac{uaf}) utilized environments engineered specifically for this research. Lacking public documentation or \ac{cve}s, latent memory retrieval was mathematically impossible. The autonomous exfiltration, Digital Twin instantiation, heap reverse-engineering, and dynamic offset calculation for complex Ret2Libc and \ac{rop} chains irrefutably prove active procedural reasoning.
    \item \textbf{Transparent Contamination as \ac{osint}:} In Scenario C (MrRobot), the agent bypassed authentication by injecting a ``Lore-Based'' dictionary (e.g., \texttt{ER28-0652}) derived from pre-trained knowledge. Rather than invalidating the architecture, this empirically simulates a human pentester utilizing \ac{osint} and contextual intuition to forge targeted attacks.
    \item \textbf{Isomorphic Abstraction Necessity:} Even with theoretical knowledge of public flaws (e.g., Scenario F, Brainpan), relying on memorized static offsets guarantees \ac{dos} due to environmental entropy. Success here stemmed exclusively from the Digital Twin's forensic container, which dynamically extracted the exact empirical Ground Truth (524-byte padding, specific \texttt{JMP ESP} gadget) to authorize a safe One-Shot execution.
    \item \textbf{Dynamic Problem-Solving:} Scenarios D (DC-2) and E (Metasploitable 3) required navigating internal HTTP redirects, bypassing virtual routing, and parsing unexpected DOM structures. Resolving these real-time environmental frictions cannot rely on static write-ups, confirming authentic ``Live'' error-recovery capabilities.
\end{itemize}

Ultimately, while benefiting from pre-trained knowledge, the Digital Twin Safety Layer and Self-Healing loop elevate the system from a simple script-reciter to a procedural analyst, assuring operational efficacy regardless of vulnerability novelty.

\subsubsection{Evaluating Operational Safety vs. Traditional Benchmarking}
While empirical evaluation of autonomous agents often relies on large-scale statistical validation, the architectural design of \textit{Automation-Exploit} prioritizes depth of exploitation over breadth of discovery. The inherently high Time-to-Compromise (TTC) is primarily driven by the I/O-bound, sequential nature of Phase 1 (Reconnaissance and Exfiltration), which accounts for over 77\% of the total operational time, rather than the Digital Twin instantiation. This sequential abstraction leap across multiple ports shifts the evaluative focus: exhaustive statistical runs become less relevant than ensuring deterministic safety on single, high-stakes targets. Furthermore, such runs are unrepresentative of targeted, ``low-and-slow'' operations.

Additionally, a direct quantitative benchmarking (e.g., comparing Success Rates or TTC) against existing frameworks introduces severe architectural asymmetries that fail to capture the primary metric of operational safety. Direct empirical benchmarking against commercial CTEM platforms (e.g., \textit{Pentera} \cite{pentera}, \textit{NodeZero} \cite{nodezero}) was precluded by their proprietary, closed-source nature, which prevents controlled testing in reproducible, air-gapped environments. Conversely, traditional Automatic Exploit Generation (AEG) systems \cite{mayhem, s2e, angr} fundamentally lack the web-navigation capabilities required to resolve mixed-protocol scenarios (Scenarios A-E).

Crucially, direct quantitative comparison with tool-augmented \ac{llm} wrappers (e.g., \textit{PentestGPT}~\cite{pentestgpt2024}) is methodologically unsound, as discussed in Section~\ref{subsec:comparative}: their \ac{hitl} paradigm conflates human operator speed with autonomous efficiency, and their structural confinement to non-destructive web vulnerabilities renders memory-corruption benchmarking operationally meaningless. Therefore, the framework's evaluation focuses on deterministic convergence and architectural isomorphism rather than purely statistical comparisons against functionally dissimilar baselines.

\subsubsection{Longitudinal Resilience and Reproducibility Dynamics}
To validate the architectural robustness against continuous updates in Cloud LLM safety alignments (RLHF) \cite{jailbreak2023}, as well as to empirically assess the framework's operational reproducibility, a targeted subset of the test suite, specifically, Scenarios C (MrRobot), F (Brainpan), and G (Custom Zero-Day), was subjected to multiple execution cycles over a staggered timeline. 

Given the inherently high Time-to-Compromise (TTC) and the strict API budget constraints associated with frontier models, performing exhaustive, large-scale statistical runs across the entire suite was computationally prohibitive. Instead, for these three representative scenarios, we established a robust baseline by conducting an initial execution (Day 0), followed by two additional, independent runs after a 30-day interval. This distribution successfully validated the system's reproducibility despite the intrinsic non-determinism of generative agents, demonstrating stable convergence across the three runs.

Concurrently, this longitudinal timeline stress-tested the framework against newly introduced exploitation filters. During the delayed executions, interactions with newly updated, state-of-the-art frontier models (e.g., Gemini 3 Pro) revealed dynamic friction: the model actively triggered safety filters during the strategic planning phase, outputting ethical refusals and refusing to generate functional code, while still providing the logical
offsets required for exploitation, as empirically observed during longitudinal evaluation.

Crucially, the Multi-Agent System (MAS) architecture absorbed this friction seamlessly. Because semantic reasoning (\textit{Accumulator}) is strictly decoupled from code generation (\textit{Drafter/Fixer}), the system autonomously extracted the raw offsets from the censored response. The ``Adversarial Hand-off'' mechanism then successfully reconstructed the Return-Oriented Programming (ROP) chains using localized models, without requiring any manual prompt adjustments. Parallel tests using baseline models (e.g., Gemini 2.5 Pro) proceeded without triggering these specific heuristic blocks. This divergence empirically proves that the framework's cognitive segregation transcends the dynamic censorship of any single underlying LLM \cite{llmagentsurvey2024}, ensuring both long-term operational resilience and statistical reproducibility.

\textbf{Tactical Adaptability and Structural Decoupling.}
The effectiveness of the \textit{Automation-Exploit} framework against evolving Cloud security measures relies on a rigorous separation of concerns between the underlying delivery infrastructure and the specific evasion strategy. In this architectural paradigm, the \textit{Adversarial Hand-off} mechanism and the local model loop serve as the persistent, invariant ``weaponry system'', that is, the structural engine capable of generating and processing offensive logic. Conversely, the adversarial templates cataloged in the external payload repository function as an interchangeable set of ``tactical payloads'' \cite{advpoetry2025}.

This decoupling ensures that the framework's core logic remains untouched even as commercial providers continuously update their \ac{rlhf} safety alignments \cite{semanticmask2025}. To empirically demonstrate this agility, a targeted double re-test was conducted on the Windows Native \ac{uaf} environment (Scenario H). While the baseline Pashto-based cross-lingual prompt was utilized as the primary payload for all standard evaluations \cite{yong2023lowresource}, its manual rotation within the generation module with a Base64-obfuscated variant yielded identical success rates by bypassing deterministic guardrails \cite{yuan2023cipherchat}. This demonstrates that while defensive filters are dynamic, the operator can maintain operational continuity by manually updating the generative instructions via the adversarial playbook, ensuring the framework adapts to new safety heuristics without requiring any systemic reprogramming of the core codebase.

\section{Ethical Considerations and Dual-Use Implications}
\label{sec:ethics}
The development of fully autonomous offensive security frameworks inherently introduces ``Dual-Use'' concerns. While \textit{Automation-Exploit} is designed to assist security teams in scaling deep penetration testing, autonomous exploitation capabilities could theoretically be repurposed by malicious actors.

To ensure ethical integrity and mitigate the risk of immediate real-world weaponization, this research relies on both procedural boundaries and current architectural limitations. Procedurally, all empirical evaluations were strictly confined to explicitly authorized, air-gapped virtual environments \cite{cyberrange2020}. Furthermore, the ``Adversarial Hand-off'' mechanism---used to bypass commercial Cloud \ac{llm} safety alignments \cite{jailbreak2023}---is presented purely as a diagnostic methodology to highlight the fragility of current generative guardrails, rather than as a malicious blueprint.

Architecturally, the framework currently acts as a natural safeguard against trivial abuse in production environments. As discussed in Section \ref{subsec:limitations}, the aggressive autonomous exfiltration phase is highly noisy. The system lacks the stealth and evasive modules required to bypass modern perimetric defenses, such as Web Application Firewalls (\acs{waf}), Intrusion Prevention Systems (\acs{ips}), and active Endpoint Detection and Response (\acs{edr}) monitoring. 

Consequently, deploying this framework against hardened, real-world infrastructures would result in immediate detection and neutralization. By openly discussing these capabilities and limitations, we adhere to responsible disclosure principles. The primary goal of this research is not to distribute a stealthy cyber-weapon, but rather to advance the scientific understanding of autonomous offensive behaviors. This transparency is crucial for empowering the defensive community to develop resilient countermeasures---such as the proposed Digital Twin Safety Layer---before threat actors fully operationalize evasive \ac{llm} agents.

\section{Conclusions and Future Work}
\label{sec:conclusions}
This paper presented Automation-Exploit, a fully autonomous multi-agent LLM framework that bridges semantic web reconnaissance and precise binary exploitation. By combining autonomous exfiltration, adversarial hand-off, adaptive pruning, and a conditional Digital Twin Safety Layer, it closes the identified research gap and enables risk-mitigated, one-shot memory-corruption exploits. Evaluation across eight scenarios, including undocumented zero-day environments, shows strong autonomous reasoning, operational efficiency, and risk-mitigated compromise of actual targets while preventing denial of service.

\subsection{Summary of Contributions}
\label{subsec:summary}

The primary contribution of this research is resolving the semantic and operational barriers historically isolating web reconnaissance from binary exploitation. As empirically validated, \textit{Automation-Exploit} bridges this ``abstraction leap''. By autonomously executing boundary-aware exfiltration to extract target executables and contextual intelligence (e.g., configuration files, source code), the system dynamically correlates exfiltrated artifacts to forge complex attack chains against adjacent services across distinct protocols.

Secondly, this work advances offensive operational safety. While Digital Twins are recognized in security analytics \cite{cyberrange2020, digitaltwin2023}, we pioneer their dynamic integration into an autonomous LLM-driven Multi-Agent architecture specifically for memory-corruption exploitation. Enforcing extreme isomorphic synchronization, i.e., \texttt{libc} alignment and runtime File Descriptor Hooking via \texttt{LD\_PRELOAD} and \texttt{dup2} \cite{kerrisk2010}, guarantees a highly accurate behavioral replica. Consequently, the autonomous agents can safely debug destructive payloads within an isolated container, enabling mathematically stabilized, isolated ``One-Shot'' executions that drastically mitigate the risk of unintended \acf{dos} on actual targets.

Furthermore, the \acf{mas} architecture introduces structural solutions to inherent \ac{llm} limitations. The hybrid ``Adversarial Hand-off'' successfully addresses commercial Safety Alignment constraints \cite{jailbreak2023} through generative decoupling, while the Decision Engine and Navigator agent's runtime Adaptive Pruning drastically optimize computational budgets. Crucially, the Two-Stage Adversarial Auditing protocol significantly reduces false positives stemming from generative hallucinations \cite{llmhallucination2025} by strictly cross-referencing payload executions with deterministic forensic telemetry, ensuring only factually verified compromises are reported.

Ultimately, the framework delivers fully autonomous, end-to-end infrastructural orchestration. Unlike solutions requiring human-in-the-loop assistance for environment setup or binary execution, \textit{Automation-Exploit} natively manages Docker hypervisors, isolated container lifecycles, and external deterministic tools. This seamless integration of abstract \ac{llm} reasoning with low-level infrastructural control represents a significant advancement in the operational independence of offensive security frameworks.

\subsection{Limitations}
\label{subsec:limitations}
Despite empirical success, \textit{Automation-Exploit} operates under specific constraints. First, architectural overhead, encompassing Digital Twin instantiation, forensic extraction, and \ac{llm} inference, yields a structurally high \acf{ttc}, restricting the framework to targeted, ``low-and-slow'' engagements rather than rapid, large-scale vulnerability assessments. Secondly, the framework's evaluation was intentionally confined to controlled, air-gapped environments. This methodological choice was made to isolate and measure the baseline cognitive capabilities of the \ac{llm} (e.g., logical reasoning, syntax generation, and self-healing) without the confounding variables introduced by active perimetric defenses. Consequently, \textit{Automation-Exploit} is currently defenseless against modern Web Application Firewalls (\acs{waf}) or Intrusion Prevention Systems (\acs{ips}). The aggressive autonomous exfiltration phase generates significant network noise, lacking the stealthiness required to evade a modern Security Operations Center (SOC). Beyond this, the framework operates strictly as an ``Initial Access'' agent mapping an external IP to a local compromise. It currently lacks the architectural modules required for post-exploitation network pivoting, lateral movement, or multi-host Active Directory enumeration, effectively restricting its use to single-node, deep-exploitation scenarios. Moreover, the Two-Stage Adversarial Auditing protocol inherently introduces a risk of circular dependency. By employing an LLM (i.e., the Reviewer) to validate the output of another LLM (the Drafter and Fixer), the system operates without a deterministic ground-truth oracle. While cross-referencing payload executions with deterministic forensic logs (e.g., \texttt{dmesg}, HTTP status codes) heavily mitigates this risk, the possibility of ``shared hallucination'', where the Auditor falsely agrees with a hallucinated log interpretation, remains a structural limitation of multi-agent LLM debate models \cite{multiagentdebate2023}. Finally, \ac{llm} reliance introduces cognitive frictions. Strict ``expert role-playing'' prompts induce an ``over-engineering'' bias, occasionally overcomplicating payloads for trivial flaws \cite{llmcontext2024, llmagentsurvey2024}. Furthermore, the system exhibits ``procedural stubbornness'' against ambiguous feedback, exhausting iteration budgets on futile syntax variations until forcibly aborted by the Auditor, which highlights the \ac{llm}'s difficulty in distinguishing correctable errors from hard defensive blocks \cite{pentestagent2025}.

Additionally, the longitudinal evaluation was conducted over three independent
runs per scenario, a choice driven by the substantial cost and duration of
each trial: each run involves multi-hour agentic sessions with sustained
interaction with commercial \ac{llm} APIs (e.g., Gemini~2.5 Pro,
Gemini~3 Pro), resulting in non-trivial per-trial expenditure~\cite{frugalgpt2023}.
This constraint is consistent with resource-limited evaluation practice
in \ac{llm}-based systems research. The low inter-run variance observed
across all eight scenarios (cf.\ Table~\ref{tab:kpi_summary}) provides
empirical evidence of result stability despite the limited repetition count.

\subsection{Future Research Directions}
\label{subsec:future_work}
To address operational constraints, future iterations will integrate asynchronous parallelization during I/O-bound reconnaissance, drastically compressing the high \acf{ttc} without impacting the \ac{llm}'s inferential budget. Additionally, transitioning to a ``Stateful Agentic Workflow'' via robust checkpointing will enable continuity across extended engagements, mirroring Advanced Persistent Threats (\acs{apt}s). 

To overcome Docker's ``Shared Kernel Limitation'', the Digital Twin Safety Layer will integrate QEMU full-system emulation, allowing the safe exploitation of kernel-space vulnerabilities (e.g., privilege escalations) to achieve absolute architectural isomorphism. Cognitively, the framework's current \textit{tabula rasa} approach will evolve by integrating ``Reflexion'' architectures and episodic memory \cite{reflexion2023}, enabling the agent to formulate persistent heuristic rules via Verbal Reinforcement Learning. Finally, regarding perimetric defenses (\acs{waf}, \acs{ips}), future research will pivot towards \textit{Adversarial Emulation}. Rather than developing stealthy modules for undetected real-world exploitation, the framework will be utilized to generate evasive, polymorphic payloads via In-Context Learning \cite{icajailbreak2024} strictly within monitored cyber ranges. This controlled generation will serve to stress-test and harden modern Security Operations Centers (SOCs) against emerging AI-driven threats, ensuring that defensive mechanisms evolve concurrently with autonomous offensive capabilities.

\section*{Availability of Data and Materials}
The custom vulnerable environments, demonstration videos, and raw forensic logs supporting the findings of this study have been submitted to the journal for peer review. To prevent misuse or unauthorized replication prior to formal publication, the supplementary repository is temporarily restricted. Demonstration videos highlighting the framework's capabilities are available via the lead author's professional channels.

\bibliographystyle{unsrt}  
\bibliography{mybibfile}

\end{document}